\RequirePackage{rotating}
\documentclass[fleqn,usenatbib]{mnras}

\usepackage{newtxtext,newtxmath}
\usepackage[T1]{fontenc}
\usepackage{ae,aecompl}
\usepackage{ dsfont }
\usepackage[autostyle]{csquotes}

\usepackage{graphicx}	% Including figure files
\usepackage{amsmath}	% Advanced maths commands
\usepackage{amssymb}	% Extra maths symbols
\usepackage{adjustbox}
\usepackage{rotating}
\usepackage{geometry}
\usepackage{pdflscape}
\usepackage{ wasysym }
\usepackage{longtable}

\title[YSO candidates in the \textit{Gaia} DR2 x AllWISE catalogue]{Identification of Young Stellar Object candidates in the \textit{Gaia} DR2 x AllWISE catalogue with machine learning methods}

\author[G. Marton et al.]{G. Marton,$^{1}$\thanks{E-mail: marton.gabor@csfk.mta.hu}
P. \'Abrah\'am$^{1}$,
E. Szegedi-Elek$^{1}$,
J. Varga$^{1}$,
M. Kun$^{1}$,
\'A. K\'osp\'al$^{1,2}$,
\newauthor
E. Varga-Vereb\'elyi$^{1}$,
S. Hodgkin$^{3}$,
L. Szabados$^{1}$,
R. Beck$^{4,1}$,
and Cs. Kiss$^{1}$
\\
$^{1}$Konkoly Observatory, Research Centre for Astronomy and Earth Sciences, Hungarian Academy of Sciences,\\
Konkoly Thege Mikl\'os \'ut 15-17, H-1121 Budapest, Hungary\\
$^{2}$Max Planck Institute for Astronomy, K\"onigstuhl 17, D-69117 Heidelberg, Germany\\
$^{3}$Institute of Astronomy, University of Cambridge, Madingley Road, Cambridge CB3 0HA, UK\\
$^{4}$Institute for Astronomy, University of Hawaii, 2680 Woodlawn Drive, Honolulu, HI 96822, USA
}

\date{Accepted XXX. Received YYY; in original form ZZZ}

\pubyear{2015}

\begin{document}
\label{firstpage}
\pagerange{\pageref{firstpage}--\pageref{lastpage}}
\maketitle

\begin{abstract} 
The second \textit{Gaia} Data Release (DR2) contains astrometric and photometric data for more than 1.6 billion objects with mean \textit{Gaia} $G$ magnitude $<$20.7, including many Young Stellar Objects (YSOs) in different evolutionary stages. In order to explore the YSO population of the Milky Way, we combined the \textit{Gaia} DR2 database with WISE and Planck measurements and made an all-sky probabilistic catalogue of YSOs using machine learning techniques, such as Support Vector Machines, Random Forests, or Neural Networks. Our input catalogue contains 103 million objects from the DR2xAllWISE cross-match table. We classified each object into four main classes: YSOs, extragalactic objects, main-sequence stars and evolved stars. At a 90\% probability threshold we identified 1\,129\,295 YSO candidates. To demonstrate the quality and potential of our YSO catalogue, here we present two applications of it. (1) We explore the 3D structure of the Orion A star forming complex and show that the spatial distribution of the YSOs classified by our procedure is in agreement with recent results from the literature. (2) We use our catalogue to classify published \textit{Gaia} Science Alerts. As \textit{Gaia} measures the sources at multiple epochs, it can efficiently discover transient events, including sudden brightness changes of YSOs caused by dynamic processes of their circumstellar disk. However, in many cases the physical nature of the published alert sources are not known. A cross-check with our new catalogue shows that about 30\% more of the published \textit{Gaia} alerts can most likely be attributed to YSO activity. The catalogue can be also useful to identify YSOs among future \textit{Gaia} alerts.

\end{abstract}


%%%%%%%%%%%%%%%%%%%%%%%%%%%%%%%%%%%%%%%%%%%%%%%%%%

%%%%%%%%%%%%%%%%% BODY OF PAPER %%%%%%%%%%%%%%%%%%

\section{Introduction}\label{intro}
Astronomy now is facing a major data avalanche. The analysis of huge amounts of data is a new challenge that researchers have to cope with. The term \enquote{data mining} is more and more widespread in the astronomical community. What is discovered with data mining and knowledge discovery in databases, can be further studied on a statistical basis. Already by April, 2019, more than 1\,300 refereed papers are available via the NASA ADS database with titles including the phrase \textquote{machine learning} in various topics like galaxy cluster mass estimation \citep[e.g.][]{armitage2019}, exoplanet transit detection \citep[e.g.][]{schanche2019}, power spectra prediction from reionization simulations \citep[e.g.][]{jennings2019} or morphological classification of 14\,245 radio AGNs \citep[e.g.][]{ma2019}. Supervised and unsupervised machine learning techniques are available to deal with problems like clustering, classification, regression and outlier detection, and they can easily outperform traditional methods, e.g. star-galaxy separation on colour--brightness and colour--colour diagrams \citep[e.g.][]{malek2013}.

One important research area, where this new attitude will bring fundamental new results, is star formation. How the Sun -- and in general, stars -- were born, is identified as one of the top four areas of modern astronomy in the European Astronet Science Vision {\href{http://www.astronet-eu.org/FP6/astronet/www.astronet-eu.org/spipca9c.html?article40}{document}\footnote{Panel C: What is the origin and fate of stars and planetary systems?}. Star formation starts as the gravitational collapse of dense, rotating, magnetic molecular cloud cores. Due to angular momentum conservation, this process is always accompanied by the formation of circumstellar disks \citep[e.g.][]{shu1987}, sometimes still embedded in their envelope. Disk-bearing YSOs can be recognized by the infrared excess emission originating from the dusty component of these structures. Another signature of newly born stars with protoplanetary disks is their brightness variations that provide information on the transitions between protostellar evolutionary stages, on the structure of their circumstellar disks, and on the different ways that protostars may affect the initial conditions of planet formation \citep[e.g.][]{abraham2009,kospal2012}. 

A long--standing problem of current star formation theories is the luminosity problem, meaning that the expected cloud core infall rates for Class I YSOs is the order of 10$^{-6}$ M$_{\odot}$ yr$^{-1}$, that should cause 10--100 times higher luminosity than observed \citep{kenyon1990, enoch2009, dunham2014}. The other problem is that the Hertzsprung-Russel diagrams (HRD) of pre--main sequence (PMS) star clusters have a wide-scatter about the best--fitting isochrone. A possible single explanation for both these problems is episodic accretion \citep[e.g.][]{aduard2014}. The star normally collects material from the surrounding envelope and disk at a low rate ($10^{-7} \mathrm{M}_\odot \mathrm{yr}^{-1}$), and at rare occasions the accretion rate rapidly increases ($10^{-4} \mathrm{M}_\odot \mathrm{yr}^{-1}$). If most of the PMS stars show some eruptive behavior then the lower--than--expected luminosity is explained by the very low accretion rate which is happening for most of the time in this evolutionary phase. On the other hand the scatter would be the result of the more dramatic accretion events, which affects the locations of the stars in the HRD \citep[e.g.][]{baraffe2009}.

\citet{lucas2017} propose that low level variations of the accretion are interpreted as an effect of the stellar magnetosphere, generating multiple, often unstable accretion flows \citep[e.g.][]{romanova2008}. More dramatic, significant eruptive events have been classically divided into two sub-classes \citep[see e.g.][]{herbig1989}: (1) FU Orionis-type variables (FUors) are thought to be the most dramatic examples of episodic accretion. FUors exhibit 5--6 mag optical outbursts attributed to highly enhanced accretion \citep{hartmann1996}. During these outbursts, accretion rates from the circumstellar disk onto the star are on the order of 10$^{-4}$ M$_{\astrosun}$yr$^{-1}$, three orders of magnitude higher than in quiescence. The exact physical mechanism of FUor outbursts is debated. Explanations include viscous-thermal instabilities in the disk \citep{bell1994}, a combination of gravitational and magnetorotational instabilities \citep[MRI]{armitage2001}, or accretion of clumps in a gravitationally fragmenting disk \citep{vorobyov2005, vorobyov2006, vorobyov2015}. Yet another type of theory involves a close stellar or substellar companion that perturbs the disk and triggers the onset of the enhanced accretion \citep{bonnell1992,lodato2004,nayakshin2012}. (2) As it is described by e.g. \citet{kospal2014}, EX Lup-type objects (EXors) are a group of low-mass pre-main sequence stars, repetitively showing  optical outbursts of  $\sim$1--5 mag.  The duration of these eruptive events covers timescales ranging from a  few  months to a few years. The outburst is  usually attributed to  enhanced accretion from the inner circumstellar disk (within $\sim$0.1 AU) to the stellar photosphere, caused by an instability in the disk \citep{herbig1977, herbig2008}. In  quiescence,  EXors  typically  accrete at a rate of 10$^{-10}$ to 10$^{-7}$M$_{\astrosun}$yr$^{-1}$, while in outburst, accretion rates are about an order of magnitude higher \citep{lorenzetti2012}. These brief  phenomena  may  significantly contribute to  the  build-up  of  the  final  stellar  mass.  Moreover, the outbursts have an effect on the circumstellar material, as well, as demonstrated by \citet{abraham2009}. They discovered episodic crystallization of silicate grains in the disk surface due to the increased luminosity and temperature during the 2008 outburst of EX Lup, resulting in material that forms the building blocks of primitive comets.

One way to resolve the above problems is if we become more efficient in identifying the eruptive phenomena of YSOs and we are able to monitor more of them with high temporal resolution, so more observational data for theories would be collected.

Variability in YSOs occurs on timescales spanning a wide range and time-domain observation of the sky is a rapidly developing field of research. Recent surveys, e.g. ASAS \citep{pojmanski1997} PTF \citep{la$W2$009}, YSOVAR \citep{morales2011}, CSI 2264 \citep{cody2014}, ASAS-SN \citep{shappee2014, kochanek2017} provided light curves of millions of sources.

Present and future observations involve at least an order of magnitude more, billions of sources, with hundreds of measured attributes per source. \textit{Gaia} \citep{gaia2016a} is a currently operational cornerstone mission of the European Space Agency. The mission is not only the most ambitious stellar astrometric project ever, but also the best transient discovery machine today. It collects photometric observations of $\sim$1.6 billion stars down to an extreme faint limit of 20.7 mag in the $G$ band and obtains low-resolution spectroscopy down to $\sim$ 19 mag for an average of $\sim$80 epochs during the nominal five year mission, although the cadence is highly depending on the scanning law \citep{wyrzykowski2012}. In the future, LSST \citep{ivezic2008} will produce 30 terabytes of data in one night and will detect stars brighter than 24.5 magnitude. The cadence of the LSST is planned to be 2--3 days (split between different filters).

Based on the \textit{Gaia} data, stellar light curves with $\sim$30 days cadence can be constructed, and analysed for variability. The task of the \textit{Gaia} Photometric Science Alerts System\footnote{http://gsaweb.ast.cam.ac.uk/alerts/home} (operated by the Institute of Astronomy at Cambridge University) is to check all light curves, and search for unpredictable transients which may herald important astrophysical events, like stellar explosions. Currently $\sim$6 alerts per day are selected from about 10$^5$--10$^6$ candidates, and posted on a dedicated \href{http://gsaweb.ast.cam.ac.uk/alerts/alertsindex}{webpage} a few days after the detection. 

Up to now, more than 7\,500 alerts were published, triggering follow-up observations and scientific papers \citep[e.g.][]{hillenbrand2018}. From this sample several thousands alerts were published as unknown type of object and only $\sim$120 were proposed to be variable YSOs by the alerts team. The latter number is astonishingly low, because decades of groundbased monitoring observations demonstrated that young stars also frequently exhibit spectacular variations, both brightenings and fadings.  Since \textit{Gaia} is the only all-sky transient survey today, a number of events would go unnoticed without the alerts. With a more efficient identification of YSOs among the alerted objects, \textit{Gaia} could make a strong scientific contribution in the field of star and planet formation, by discovering and publishing otherwise unnoticed YSO brightenings and fadings.

In this paper we aim at identifying YSO candidates based on the second data release of the \textit{Gaia} mission. Probability based classification of $\sim$103 million sources into four main object classes (Y -- YSO candidates, EG -- extragalactic sources, MS -- main sequence stars, E -- evolved stars) was done using multi-band photometric data. Section~\ref{dataandmethods} describes the catalogues and maps we used in our study, also gives a brief summary of the different classification methods. In Section~\ref{results} we describe how effectively the different methods performed in our tests. In Section~\ref{discussion} we apply our YSO candidate selection to a star forming cloud and compare our results to that of a recently published study. We also cross-match our candidates with the published and unpublished \textit{Gaia} alerts, analyse their flux asymmetry distribution and find that we can increase the number of YSO related alerts among the studied sample by 50\%.

\section{Data and methods}\label{dataandmethods}
\subsection{Catalogues and maps}
\subsubsection{\textit{Gaia} DR2}
The second release of the \textit{Gaia} Data (DR2) was made available on 2018 April 25$^{th}$, consisting of astrometry and photometry for over 1.6 billion sources brighter than magnitude 20.7 in the $G$ band \citep{gaia2018}. \textit{Gaia} DR2 is based on observations collected between 2014 July 25$^{th}$ and 2016 May 23$^{rd}$. The main content of the catalogue and the limitations of the DR2 are well explained on the \href{https://www.cosmos.esa.int/web/gaia/dr2}{Gaia Cosmos Webpage} and in \citet{gaia2018}.

For our science case it is important to note that while the DR2 catalogue is essentially complete between $G=12$ and $G=17$, it has an ill-defined faint magnitude limit, which depends on celestial position. YSOs, especially at the early stages of their evolution, not only emit the majority of their radiation at near- and mid-infrared wavelengths and are fainter in the part of the electromagnetic spectrum visible to $Gaia$, but being typically associated with dense clouds and accreting envelopes and disks they suffer larger extinctions than main sequence and evolved stars. Based on our training samples (see Section~\ref{section:trainingsamples}) we found that 55\% of the YSOs detected by the $Spitzer$ Space Telescope \citep{werner2004} is present in the DR2 and 68\% of those are brighter than 17 magnitudes in the $G$ band. Because of the scan law pattern, source density fluctuations also exist, which affect the completeness of crowded regions, like star forming regions. Crowding and blending are important features of star forming regions, and especially at fainter magnitudes ($G>19$) the photometric measurements from the blue and red photometers suffer from an insufficiently accurate background estimation and from the lack of specific treatment of blending and contamination from nearby sources.

\subsubsection{AllWISE and 2MASS catalogue}
\textit{WISE} is a NASA space telescope that was launched in  December  2009.  It scanned the whole sky in four near-- and mid--infrared passbands: $W1$, $W2$, $W3$ and $W4$, centered at at  3.4,  4.6,  12,  and  23 $\mu$m, respectively. The AllWISE Source Catalogue \citep{cutri2013} was produced by combining  the  \textit{WISE}  single-exposure  images  from  several survey phases and contains 747\,634\,026 sources. According to the \href{http://wise2.ipac.caltech.edu/docs/release/allwise/expsup/sec2\_3a.html}{explanatory supplement} the angular resolution is 6$\arcsec$.1, 6$\arcsec$.4, 6$\arcsec$.5, and 12$\arcsec$.0, in bandpasses W1--W4 respectively, with 5$\sigma$ detection limit estimated to be 0.054, 0.071, 0.73, and 5 mJy. 

%YSOs are embedded and/or surrounded by dense circumstellar material, therefore they are less visible in the optical, but the thermal emission of the heated dust particles peaks at near-- and mid--infrared wavelengths. 

The \textit{WISE} telescope \citep{wright2010} was not designed for observation of faint and deeply embedded objects, but still provided the most recent all-sky survey in the part of the electromagnetic spectrum, where YSOs usually emit the bulk of their radiation.

The AllWISE catalogue already includes the $J$, $H$ and $K_s$ band photometric data from the 2MASS point- and extended source catalogues \citep{skrutskie2006}. 

\subsubsection{Planck foreground map}\label{planckmap}
YSOs are located predominantly in regions containing a large amount of dust. The \textit{Planck} space telescope collected infrared light emitted by the dust particles for the all-sky. In order to decide if a source is associated with a dusty region we extracted dust opacity ($\tau$) value for each source. This value also gives us a clue about the volume of interstellar extinction which has a major impact on the colour of the sources. We used the 353 GHz R2.01 \textit{Planck} dust opacity map \citep{planck2016} value at the position of each object. The map is a HealPix image, with NSide=2048 and a pixel resolution of 1.718$\arcmin$.

\subsection{Initial dataset}\label{initialdata}
As explained in Section~\ref{intro}, the spectral energy distribution (SED) of YSOs with circumstellar disks show an infrared excess. Automated classifiers designed to identify disk bearing YSOs provide optimal performance if we include infrared data. Therefore, in this study we are only interested in those $Gaia$ DR2 objects which have matching detections in the AllWISE catalogue. The cross-match of the AllWISE and DR1 catalogues was done by \citet{marrese2017}. Their method accounts for the proper motion, different epochs and also e.g. the local surface density of the external catalogue. This exercise was repeated for the DR2 and the cross-matching table was part of the official data release \citep{marrese2018}. For our case we used the AllWISE BestNeighbour table that contained 300\,207\,917 matches. The table listed the \textit{Gaia} Source IDs and the AllWISE source IDs for all pairs.

As we are mostly interested in finding YSOs and YSO candidates, we explored only a subset of the \textit{Gaia} DR2 X AllWISE table. To identify the regions where YSOs are mostly located, we checked the \textit{Planck} dust opacity map (Section~\ref{planckmap}) values at the celestial positions of the known YSOs, collected from the SIMBAD database, the \textit{Spitzer} archives and the literature (see Section~\ref{section:trainingsamples} for more details). We found that 99\% of the known YSOs are located in regions where the dust opacity value is higher than 1.3$\times 10^{-5}$. 

The sources outside these regions, and those lacking data in one or more photometric bands were removed from our list. We also removed those objects where multiple $Gaia$ IDs were associated with one AllWISE object. As a result our initial dataset contained 101\,838\,724 sources including \textit{Gaia} optical G magnitudes, $J$, $H$, $K_s$ magnitudes from 2MASS, AllWISE magnitudes, and the optical depth ($\tau$) from the Planck dust opacity map.

\subsection{Training sample}\label{section:trainingsamples}

Finding YSO candidates among unknown sources is a typical classification problem. We used supervised machine learning techniques to identify YSO candidates. For such algorithms training samples needs to be created. A training sample contains data of objects from known classes. The problem to be solved is to find the best algorithm and those properties (i.e. colours, source extension, environmental parameters) which describe the different classes in the most meaningful way and can separate the classes the most efficiently. The goal is to maximize the number of true positive (e.g. a known YSO is classified as YSO) and minimize the number of false negative classifications (when a YSO is classified as something different). 

In our previous works we attempted to identify YSO candidates based on \textit{AKARI} data combined with \textit{WISE} photometry \citep{toth2014} and on purely AllWISE data \citep{marton2016}. In these attempts our training samples were based on the \href{http://simbad.u-strasbg.fr/simbad/}{SIMBAD database}. SIMBAD includes astronomical objects from scientific publications. It provides information on the nature of the objects via the \textit{main\_type} and \textit{other\_type} columns. The \textit{main\_type} column includes information on the most specific type of the object (if known), which is the primary object type, while the \textit{other\_type} lists the primary and the more generic types, e.g. the AllWISE source J173821.67-293459.5 has a ``YSO'' \textit{main\_type}, but the \textit{other\_type} is ``Y*O|IR''. Still, in many cases the \textit{main\_type} is a general object type, like ``Galaxy'' or ``Star''.

In this study we searched not only the SIMBAD, but also used $\sim$80 catalogues from the literature to identify the known YSOs and other object types in order to have the best possible training samples. All the objects were grouped into four object classes: Main-sequence stars (MS); Extragalactic objects (EG); Evolved stars (E); YSOs (Y). These classes were built from the following datasets and catalogues:
\begin{itemize}
\item Main-sequence stars (MS): 
	\begin{itemize}
	\item Main-sequence stars below 6500 K observed by the Kepler Space Telescope \citep{mcquillan2014}
    \item Main-sequence stars from the 3rd edition all-sky catalog of \citet{kharchenko2001}
    \item Main-sequence stars from the Catalogue of stellar UV fluxes \citep{thompson1978}
	\end{itemize}
\item Extragalactic objects (EG): 
	\begin{itemize}
	\item SIMBAD main types equal to AGN, AGN\_Candidate, BClG, BLLac, BLLac\_Candidate, Blazar, Blazar\_Candidate, BlueCompG, Compact\_Gr\_G, EmG, Galaxy, GinCl, GinGroup, GinPair, GravLensSystem, GroupG, HII\_G, IG, LensedG, PairG, PartofG, QSO, QSO\_Candidate, RadioG
	\item Objects from the Allwise AGN \textit{Gaia} DR2 cross-identification table
	\end{itemize}

\item Evolved stars (E)
	\begin{itemize}
	\item SIMBAD main types of Mira, RGB*, post-AGB*, RRLyr, AGB* and HB*
    \item Objects of the \textit{Gaia} DR2 Cepheid stars table and RR Lyrae stars table
	\end{itemize}
\item YSOs (Y)
	\begin{itemize}
	\item Photometric YSO catalogues found in the \href{http://vizier.u-strasbg.fr/}{VizieR database} (Appendix \ref{appendix:a})
    \item Spectroscopic YSO catalogues listed by VizieR (Appendix \ref{appendix:b})
    \item YSO related Spitzer publications (Appendix \ref{appendix:c})
    \item Spitzer YSOs and YSO candidates from the project "From Molecular Cores to Planet-Forming Disks" (C2D) \citep{evans2003}
	\end{itemize}
\end{itemize}

The objects of the above catalogues were cross-matched with our initial dataset (Section~\ref{initialdata}). In Figure~\ref{fig:sep} we show the distribution of the angular distance values between the \textit{Gaia} DR2 positions and that of the \textit{Spitzer} YSOs and YSO candidates. In 95\% of those cases where a counterpart was found within 5$^{\prime\prime}$, the matching source was closer than 1$^{\prime\prime}$. We also checked the distances between the $Gaia$ DR2 and the AllWISE counterpart of these sources, listed in the $Gaia$ DR2 x AllWISE table and found that 99.7\% of the best neighbours are within 1$^{\prime\prime}$. Thus, we decided to use a 1$^{\prime\prime}$ search radius in the procedure, although it is an arbitrary number, but it fits our requirements.

The cross-match resulted in 50\,001 EG, 182\,656 MS, 63\,372 E and 14\,302 Y class objects that were used to create the training samples. The $G$ band brightness--$\tau$ and the $G-W1$ vs. $W1-W2$ colour-colour {diagram of all these sources} are plotted in Figure~\ref{fig:bt} and Figure~\ref{fig:cc} in Appendix~\ref{appendix:diagrams}, representing how the different object types are overlapping in the 2D cuts of the multi-dimensional feature space. For each training sample 14\,000 objects of each class were selected randomly. We used equal number of elements from each class to avoid a bias caused by the order of magnitude difference between the different classes. The elements were then permuted and 7\,000 objects of each class was selected as the actual training sample, while the remaining 7\,000 were used to check the goodness of the classification. The permutation and selection was repeated 10 times, allowing us to have a statistically more meaningful classification procedure.

\begin{figure}
	\includegraphics[width=0.45\textwidth]{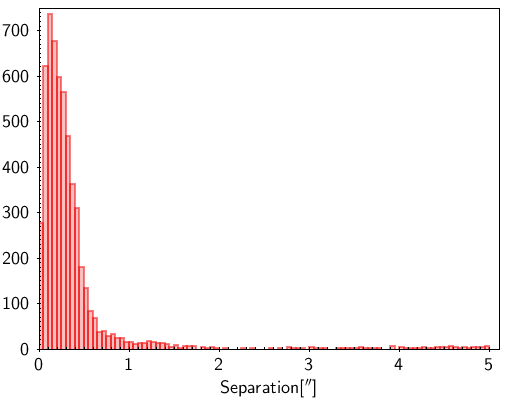}
    \caption{Distribution of angular distance values between the \textit{Gaia}DR2xAllWISE and the \textit{Spitzer} YSO catalogue positions. 95\% of the matching \textit{Spitzer} sources are within 1$^{\prime\prime}$ radius of the DR2 position.}
    \label{fig:sep}
\end{figure}

\subsection{Classifiers}
Classification with supervised methods requires three major steps:
\begin{enumerate}
\item Training the classifier
\item Classifying new data with the classifier
\item Tuning the classifier
\end{enumerate}
To do so, several statistical softwares and packages are available for use. In our work we used R (R Core Team 2013)\footnote{http://www.R-project.org}, a free software environment for statistical computing and graphics, commonly used in the astronomical community.

We experimented with several classifier methods in order to evaluate their performance and to select the one which is best for our needs. These methods are the Support Vector Machines \citep[SVM, e.g.][]{solarz2017,krakowski2016,kurcz2016,marton2016,heinis2016,kovacs2015,solarz2015,malek2013,fadely2012,beaumont2011}, the $k$--Nearest Neighbors \citep[$k$--NN, e.g.][]{pashchenko2018, lochner2016,buisson2015}, the Naive Bayes \citep[e.g.][]{mitchell1997,lochner2016}, Neural Networks \citep[NN][]{venables2002} and Random Forests \citep[RF, e.g.][]{breiman2001, buisson2015, pashchenko2018}.

\subsection{Dimension reduction}

While all machine learning techniques are capable of handling multi-dimensional datasets, and some of them are capable to assign weights to the features (i.e. the characteristics that define our problem) in order to optimize the classification process and maximize the classification goodness, one should still consider reducing the number of variables. By selecting the relevant features we can shorten the training times, simplify our models, avoid the curse of dimensionality and reduce overfitting of the data. Our initial parameter space contained 39 columns, including the $\tau$ value from the \textit{Planck} dust opacity map, the \textit{Gaia} $G$ band brightness value, the 2MASS brightness values, the \textit{WISE} brightness values, and all the possible colours that can be calculated from the \textit{Gaia}, 2MASS and \textit{WISE} photometry. 

To reduce the dimensionality of our feature space we calculated the Pearson's correlation coefficient between each feature and created a correlation plot (see Figure~\ref{fig:corr} for the YSOs). From the left to right (and top to bottom) the coefficient value was checked and in cases the correlation was above $0.7$ or below $-0.7$, the longer wavelength feature was removed (as typically the angular resolution and background contamination increases with the wavelength). For example, in the Y class training sample (Figure~\ref{fig:corr}) the correlation coefficient value between the \textit{Gaia} G mag and the 2MASS $J, H$ and $K_\mathrm{s}$ magnitudes was 0.88, 0.79 and 0.72, respectively, so all the 2MASS brightness values were removed from the list of features. This procedure was repeated for all the classes. We ended up with the following list of meaningful features for the classes:

\begin{itemize}
\item \textbf{E}: $\tau, G, H, W4, G-W4, J-H, K_\mathrm{s}-W1, W1-W2, W1-W3 $
\item \textbf{EG}: $\tau, G, J, W2, W3, G-J, G-W2, J-H, J-K_\mathrm{s}, H-W2, W2-W3, W3-W4 $
\item \textbf{MS}: $\tau, G, W4, G-J, G-W4, H-K_\mathrm{s}, H-W1, H-W3, H-W4, K_\mathrm{s}-W2, W1-W2, W1-W3 $
\item \textbf{Y}: $\tau, G, W1, G-J, G-W4, J-H, H-W1, W1-W2, W1-W3, W2-W4, W3-W4 $
\end{itemize}

Finally, the union of the above features was used in the training sets, resulting in 25 columns instead of the 39 used initially: $\tau, G, J, H, W1, W2, W3, W4, G-J, G-W2, G-W4, J-H, J-K_\mathrm{s}, H-K_\mathrm{s}, H-W1, H-W2, H-W3, H-W4, K_\mathrm{s}-W1, K_\mathrm{s}-W2, W1-W2, W1-W3, W2-W3, W2-W4, W3-W4$. 

\begin{figure}
\includegraphics[width=0.45\textwidth]{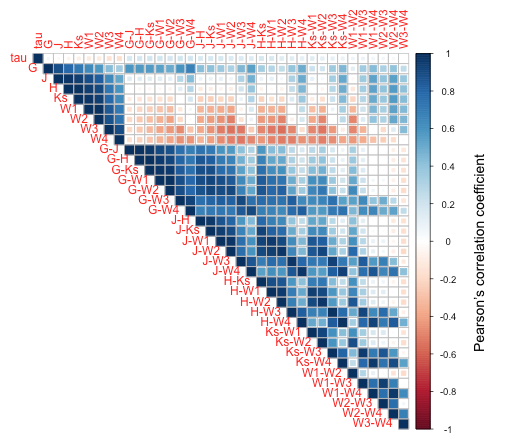}
\caption{An example of the correlation matrices of the feature space, shown for the Y class (YSOs). Strong correlation is shown with dark blue, strong anti-correlation is shown with dark red colours.}
\label{fig:corr}
\end{figure}

\subsection{Metrics}\label{section:bestclassifier}
To interpret the results of the different classifiers one has to define metrics that characterizes them in a uniform way. If the predicted class of an object is the same as the true class,  i.e. a known YSO is classified as a YSO, we call it a true positive (TP) classification. If the same known YSO is classified as another type of object, then it is a false negative (FN) classification. If an object of an extragalactic nature is classified as a YSO, it is a false positive (FP) classification.

Completeness is the percentage of known YSOs classified as YSO ($N_\mathrm{TP}/N$, where N is the number of all objects in the training set), which in ideal case is 100\%. In the best case the purity ($N_\mathrm{TP}/(N_\mathrm{TP}+N_\mathrm{FP})$) is also 100\%, meaning that all objects from a given object class is classified into their true class, e.g. all known YSOs are classified as YSO, and nothing else is classified as YSO. The contamination is the fraction of false positives among the objects classified as YSO ($N_\mathrm{FP}/(N_\mathrm{TP}+N_\mathrm{FP})$), which is 0\% if the classification is perfect.

\subsection{Identification of spurious $W3$ and $W4$ AllWISE photometry}\label{spurious}
The main science goals of the $WISE$ mission were to study infrared-bright galaxies, to find brown dwarfs, and to study near-Earth asteroids. The method of source identification on $WISE$ images is described in the online explanatory supplement\footnote{http://wise2.ipac.caltech.edu/docs/release/allwise/expsup/} and in \citet{marsh2012}. Unfortunately, this method leads to many spurious photometric results in the Galactic plane, which was already investigated in numerous papers  \citep[e.g.][]{koenig2014, marton2016, silverberg2018}. To avoid false classification by the $W3$ and $W4$ band photometry we carried out an analysis similar to what we presented in \citet{marton2016} based on the findings of \citet{koenig2014}.

\citet{koenig2014} found that the spurious and real detections in the $W3$ and $W4$ bands can be separated by using the following parameters, provided by the AllWISE catalogue:
\begin{itemize}
    \item Signal--to--noise ratio (SNR) calculated in the $W3$ and $W4$ bands
    \item The reduced chi-square parameter in the same bands
    \item The w3m, w4m parameters (the number of individual 8.8s exposures on which a profile-fit measurement of the source was possible)
    \item The w3nm and w4nm parameters (the number of individual 8.8s exposures on which this source was detected with SNR$>$3)
\end{itemize}

\citet{koenig2014} plotted these parameters and defined a set of criteria to decide if the detection can be accepted as real or if it is rejected and labeled as spurious detection. Instead of defining a set of criteria, we used the Random Forest method to classify the sources into the two classes, but we used the same parameters to separate spurious and real detections in the $W3$ and $W4$ bands.

The training sample for this exercise was created by visually inspecting the $WISE\ W3$ and $W4$ images at the positions of known YSOs. We selected 500 cases where the sources were obviously real and 500 where we could not clearly identify a source. 10 examples for both real and fake sources are presented in Figure~\ref{fig:sources}.

\begin{figure}
	\includegraphics[width=0.5\textwidth]{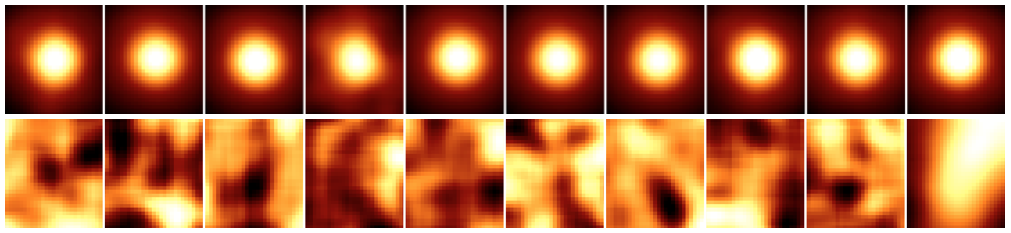}
    \caption{\textbf{Top row:} $WISE\ W4$ images of 10 sources accepted as real ones. \textbf{Bottom row:} $WISE\ W4$ images of 10 sources we labeled as fake sources. All images are centered on the position of known YSOs}
    \label{fig:sources}
\end{figure}

\section{Results}\label{results}
Below we detail the results of our classification methods. In Section~\ref{classification} we describe how the different methods performed on our training samples including long wavelength \textit{WISE} data. As explained in Section~\ref{spurious} the $W3$ and $W4$ band \textit{WISE} photometry cannot be trusted in every case, especially in crowded regions with infrared luminous backgrounds where YSOs are typically located. In Section~\ref{spuriousresult} we show how well our identification of the spurious \textit{WISE} photometry worked. In Section~\ref{withoutw3w4} we describe our classification results obtained without the $W3$ and $W4$ band photometry. Finally, Section~\ref{ysocandidates} describe our sample we consider as a reliable YSO candidate catalogue.

\subsection{$Gaia$ DR2 x AllWISE source classification including $W3$ and $W4$ \textit{WISE} photometry}\label{classification}
The metrics described in Section~\ref{section:bestclassifier} (completeness, purity, and contamination) were calculated for the YSOs with 13 different methods. SVM was used with 5 different kernels: linear, 2$^{nd}$ and 3$^{rd}$ order polynomial, radial and sigmoid. Naive Bayes was used with and without kernel estimation, the $k$-NN method was tried with 3 and 8 nearest neighbours, the Random Forests were used with 500 trees with 5 random variables at each split, 1000 trees with 7 random variables, and 2000 random trees with 10 random variables. Finally, the Neural Networks were also used with its default parameters. We plotted the results of each method in Figure~\ref{fig:comparison} and sorted them according to the purity. The blue bars show the completeness, the green bars show the purity and the red bars stand for the contamination rate.

\begin{figure}
	\includegraphics[width=0.5\textwidth]{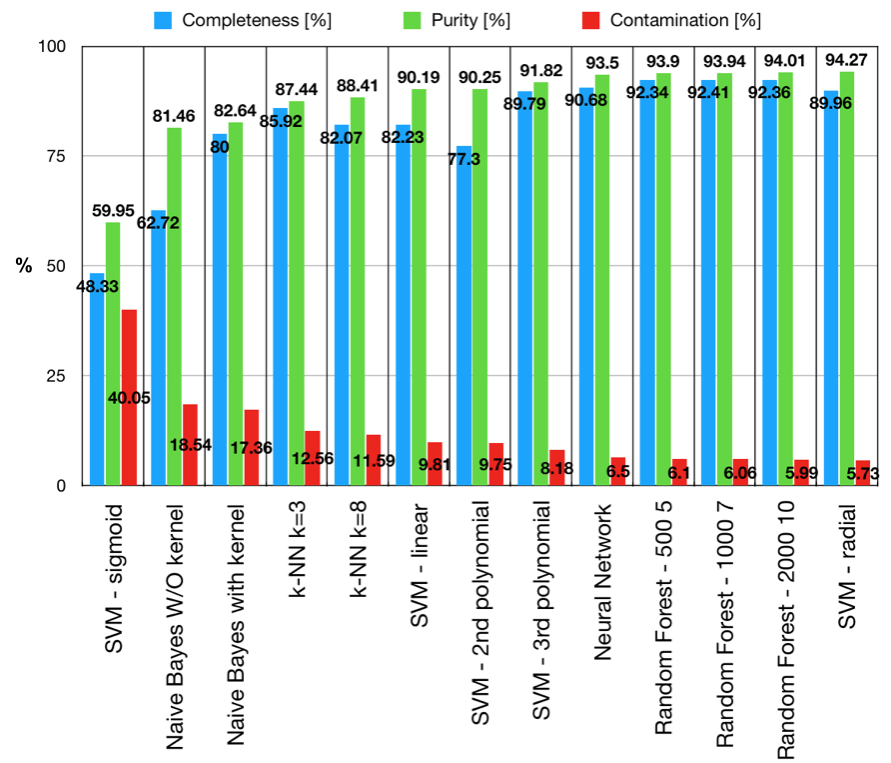}
    \caption{Completeness, purity and contamination rates achieved with different methods to classify the YSO training samples. Methods are in increasing order according to the purity (green bars) achieved with them. Blue bars indicate the completeness, red bars show the contamination rate.}
    \label{fig:comparison}
\end{figure}

Figure~\ref{fig:comparison} shows that the best classification was achieved by using the Random Forests. In all the three cases (with 500, 1000 and 2000 trees, and 5, 7 and 10 variables tried at each split, respectively) the method performed above 92\% completeness for the Y class. The Neural Network also performed above 90\%, followed by the 3$^{\mathrm{rd}}$ degree polynomial and the radial basis SVMs, close to 90\%.

Not shown in the paper, but we also compared the results achieved with the different methods using the full dataset and after dimension reduction. We found that using a reduced number of features did not improve any classification method significantly, however all of them seem to perform marginally better, except the SVM with sigmoid kernel and the Random Forest with 2000 trees, but the differences between the results is well within the estimated uncertainties. 

Our final conclusion was that the best method for our purpose is the Random Forests with 500 trees and 5 random variables tried at each split after dimension reduction. %From this point, unless it is otherwise indicated, we use sources classified as YSO candidates by this method with probability at least 0.9 ($P_Y\geq$ 0.9) and we call them YSO candidates.

On Figure~\ref{fig:bt} and Figure~\ref{fig:cc} in Appendix~\ref{appendix:diagrams} we showed how the four different object types populate the bightness-$\tau$ and colour-colour diagrams. Figure~\ref{fig:classifiedbt} and Figure~\ref{fig:classifiedcc} show the same features for the classified sources. It is apparent that the distribution of the dots is very similar in the two cases.

\subsection{Spurious photometry identification in the $W3$ and $W4$ \textit{WISE} bands}\label{spuriousresult}
As we described in Section~\ref{spurious} the $W3$ and $W4$ detections suffer from fake source identifications and spurious point source photometry. In Section~\ref{classification} we found that the Random Forests with 500 trees and 5 random variables gave us the best results, therefore for the spurious source identification was done with this method, as well.

We found that in all cases the real sources (visually identified on $W3$ and $W4$ images, see Figure~\ref{fig:sources}) were classified as real sources, so the completeness achieved in this exercise was 100\%. It also means that none of the real sources was classified as fake source. Among all our tests we found only 2 cases when a fake source was classified as real. This means that the contamination rate was as low as 0.02\% and the purity of our real source identification was 99.98\%. We note that these results were achieved by accepting that a source is classified as real, if the probability of being real ($P_R$) is higher than 0.5.

After applying our classification scheme to the initial dataset (described in Section~\ref{initialdata}) we found that 9\,816\,124 sources have $P_R > 0.5$ among the 101\,838\,724 sources. These are the sources for which we trust the $W3$ and $W4$ photometry listed in the AllWISE catalogue.

\begin{figure}
	\includegraphics[width=0.5\textwidth]{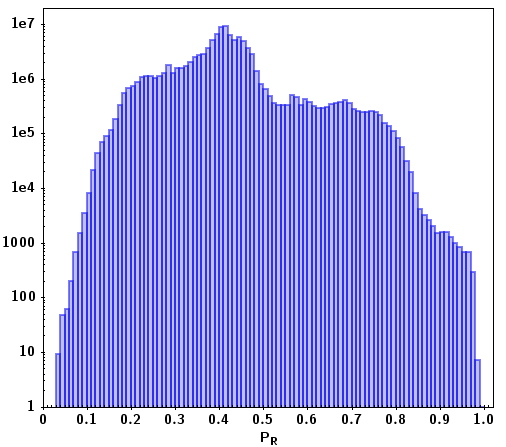}
    \caption{Distribution of probabilities of being a real source ($P_R$) among the sources of our initial dataset. 9\,816\,124 sources were classified as real sources ($P_R>0.5$), meaning that their $W3$ and $W4$ photometry can be trusted.}
    \label{fig:prdist}
\end{figure}

\subsection{$Gaia$ DR2 x AllWISE source classification excluding $W3$ and $W4$ \textit{WISE} photometry}\label{withoutw3w4}
In the previous section (Sect. \ref{spuriousresult}) we showed that only 9\,816\,124 AllWISE sources were classified as real, and the 90.4\% of the sources have spurious photometry in the longer wavelength \textit{WISE} bands. Therefore the classification into the four object classes (EG, MS, Y, E) was repeated, but without any columns containing $W3$ or $W4$ data in the training sets. Again, we used the Random Forest algorithm with 500 trees and 5 random variables at each split. The completeness was 91.78\%$\pm$0.28\%, the purity was 93.14\%$\pm$0.22\% and the contamination was 6.86\%$\pm$0.22\%.

\subsection{YSO candidates}\label{ysocandidates}

Table~\ref{tab:classificationnumbers} details the number of sources classified into each object classes depending on the adopted probability threshold of belonging to the given class. Each number is a sum of two numbers:
\begin{enumerate}
    \item The number of sources belonging to the given object class with $PL$ (classification including longer wavelength $W3$ and $W4$ \textit{WISE} data) above the threshold and $P_R$ (probability of having a real $W3$ and $W4$ photometry greater than 0.5)
    \item The number of sources belonging to the given object class with $PS$ (classification excluding longer wavelength $W3$ and $W4$ \textit{WISE} data) above the threshold and $P_R$ (probability of having a real $W3$ and $W4$ photometry lower than or equal to 0.5)
\end{enumerate}

\begin{table}
\caption{Number of sources classified as main sequence star (MS), evolved star (E), extragalactic object (EG) or YSO candidate (Y) with probabilities above the listed thresholds (0.5, 0.9, 0.95, 0.99 and 0.995). Upper table shows the numbers in the case when $W3$ and $W4$ photometry was included and $P_R>0.5$. Lower table includes numbers when classification was done without the longer wavelength \textit{WISE} photometry and $P_R\leq0.5$.}
\label{tab:classificationnumbers}
\begin{tabular}{l|cccc}
$P_R>0.5$				&	MS			&	E				&	EG				&	Y 				\\
\hline
%P$>$P$_\mathrm{others}$	&	3\,194\,834	&	64\,154\,578	&	22\,588\,740	&	13\,056\,422	\\
$P_L\geq$0.50			&	1\,017\,841	&	7\,675\,537 	&	45\,028     	&	528\,114	\\
$P_L\geq$0.90			&	458\,260	&	4\,252\,425 	&	19\,235			&	57\,710	\\
$P_L\geq$0.95			&	365\,275	&	2\,972\,483		&	16\,019			&	23\,502	\\
$P_L\geq$0.99			&	233\,647	&	888\,463		&	8\,585			&	4\,946	\\
$P_L\geq$0.995	    	&	185\,416	&	417\,842		&	5\,756			&	2\,885	\\
\hline
\hline
$P_R\leq0.5$			&	MS			&	E				&	EG				&	Y 				\\
\hline
%P$>$P$_\mathrm{others}$	&	3\,194\,834	&	64\,154\,578	&	22\,588\,740	&	13\,056\,422	\\
$P_S\geq$0.50			&	1\,596\,898	&	47\,511\,206	&	14\,500\,223	&	12\,528\,308	\\
$P_S\geq$0.90			&	62\,710 	&	8\,224\,488 	&	59\,375			&	1\,710\,918	\\
$P_S\geq$0.95			&	35\,245 	&	3\,544\,466		&	40\,068			&	235\,995	\\
$P_S\geq$0.99			&	14\,277 	&	430\,306		&	19\,706			&	15\,942	\\
$P_S\geq$0.995	    	&	8\,219  	&	169\,504		&	12\,683			&	7\,408	\\
\hline
\end{tabular}
\end{table}

We found 1\,768\,628 sources that were classified as potential YSO candidates, meaning that $P_R>0.5$ and $PL_Y\geq0.9$ or $P_R\leq0.5$ and $PS_Y\geq0.9$. Our results are published in a form of a VizieR table. The table contains the \textit{Gaia} source IDs, the AllWISE source designations, the \textit{Gaia}, 2MASS and \textit{WISE} photometric data. For each source the membership probabilities are also listed in the table. These probability values are the $P_R$ values (probability of having a real $W3$ and $W4$ photometry), the classification probabilities using all $WISE$ bands ($PL_{MS}$, $PL_E$, $PL_{EG}$, $PL_Y$) and the classification probabilities including only the shorter wavelength $WISE$ bands ($PS_{MS}$, $PS_E$, $PS_{EG}$, $PS_Y$). Because all these data are provided, users can decide about the thresholds for themselves according to the needs of their scientific study. 

%Among these candidates 360\,954 have $P_Y\geq 0.95$ and 16\,163 have $P_Y\geq 0.99$. For each source of our initial dataset (see Section~\ref{initialdata}) the probabilities are available via the VizieR database. We also publish the \textit{Gaia}, 2MASS and \textit{WISE} photometric data used for the classification.

\section{Discussion}\label{discussion}

\subsection{Application of the YSO selection to Orion A}

Recently \citet{grossschedl2018} analysed the 3D structure of the archetypal molecular cloud, Orion A based on \textit{Gaia} DR2 data. Their results show that Orion A is not a straight filamentary cloud as it appears in the sky, but has a cometary-like shape. It has two components, a denser head at $\sim$400 pc, and a longer tail with fewer young stars reaching out to $\sim$470 pc. 
We repeated their analysis but instead of using the YSOs defined in their study, we used our own YSO candidates. We started with sources we classified as YSOs candidates and created a pre-selection of the sample by applying the parallax criteria of \citet{grossschedl2018} to these sources. Moreover, we restricted our analysis to sources located between 300 and 600 pc according to the distance estimate of \citet{bailer2018}. As a result we ended up with 1\,368 objects, plotted on the \textit{Herschel}/PACS images of Orion A in Figure~\ref{fig:orionayso}. 

Figure~\ref{fig:orionaysodist} shows the median distance value of the YSO candidates calculated in $0^{\circ}.5$ bins along the galactic longitude. Our results (green line) show a very good agreement with those of \citet{grossschedl2018} presented with the black line and confirm that the head part of the cloud is located at a distance of $\sim$380--400 pc while the tail part is reaching out to $\sim$430--480 pc. We obtained similar results when using all our YSO candidates between 300 and 600 pc distance without the parallax criteria (red line). As a comparison we overplotted those sources classified as main-sequence stars (blue line) and found that their distribution is different than that of the YSOs.

\begin{figure*}
	\includegraphics[width=\textwidth]{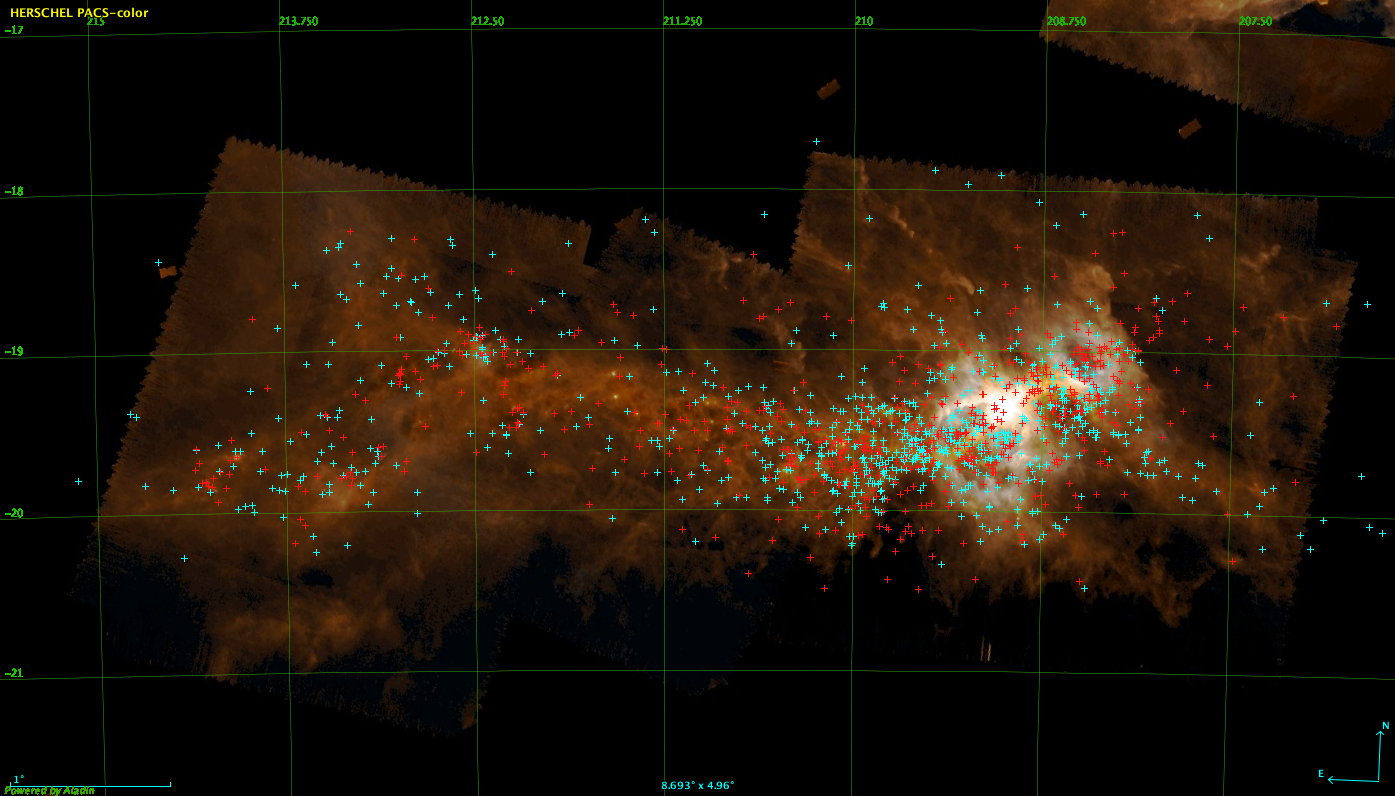}
    \caption{Our YSO candidates (cyan crosses) in Orion A overlaid on a false-color Herschel/PACS image, using galactic frame. 1\,368 YSO candidates are selected after applying the parallax criteria of \citet{grossschedl2018} and restricting their distance between 300 and 600 pc. 682 YSOs of \citet{grossschedl2018} are overplotted with red crosses.}
    \label{fig:orionayso}
\end{figure*}

\begin{figure}
	\includegraphics[width=0.5\textwidth]{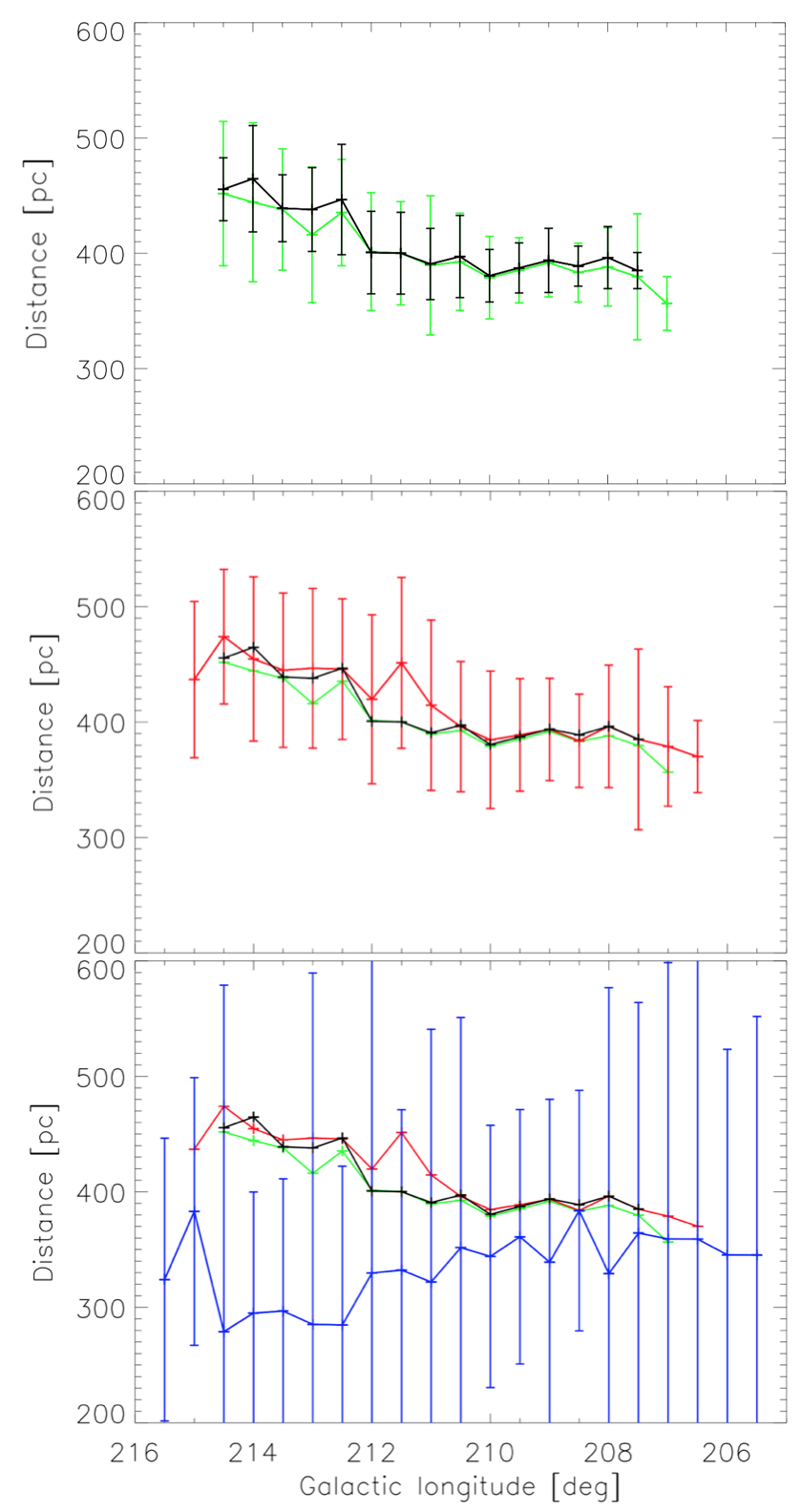}
    \caption{Median distance of the YSO candidates in Orion A as a function of the galactic longitude. The median of the distances were calculated in $0^\circ.5$ bins. The error bars represent the standard deviation of the distance values in the bins. The sample of \citet{grossschedl2018} is plotted with black. Our YSO sample with the parallax criteria of \citet{grossschedl2018} applied is shown with green. The red line present the median distance value of all of our 2\,668 YSO candidates in the region (without parallax criteria, but located between 300 and 600~pc). The blue line presents the distance distribution of the 669 sources classified as main-sequence stars (MS). They are distributed in a more homogeneous way as a function of distance, therefore the error bars are larger than for the YSOs.}
    \label{fig:orionaysodist}
\end{figure}

This field was also used to validate the purity of our selection in two ways. First, we looked for counterparts of the \citet{grossschedl2018} catalogue of \href{http://vizier.u-strasbg.fr/viz-bin/VizieR?-source=J/A+A/619/A106\&-to=3}{682 YSOs} in our initial sample, using a 1$^{\prime\prime}$ search radius. 568 matches were found. 517 of them (90\%) were classified as YSO candidate by our method. 263 of them were classified as real ($P_R>0.5$) with $PL_Y\geq0.9$, and 254 not real ($P_R\leq0.5$) with $PS_Y\geq0.9$. Figure~\ref{fig:orionaysomatch} shows the 682 \citet{grossschedl2018} sources and our YSO candidates. The distribution of the two datasets are very similar.

\begin{figure}
	\includegraphics[width=0.48\textwidth]{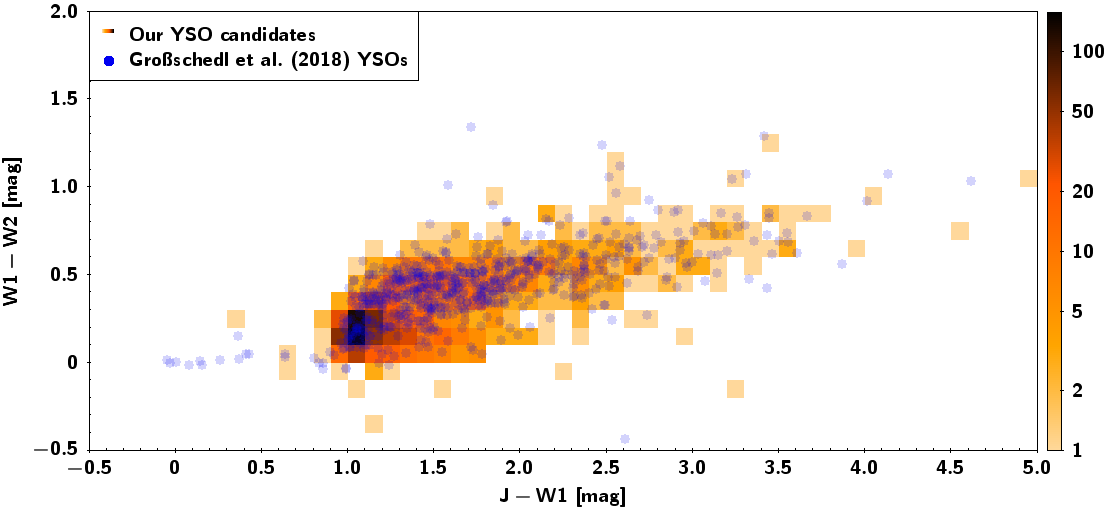}
    \caption{YSOs from the \citet{grossschedl2018} catalogue (blue opaque dots) and the surface density of our sources classified as YSO candidates (red colour coding) on the 2MASS $J$-$WISE$ $W1$ vs. $WISE$ $W1-W2$ colour-colour diagram. The surface density was calculated in bin size of 0.1 in both directions.}
    \label{fig:orionaysomatch}
\end{figure}

As an additional sanity test of our YSO classification, we cross-matched the 2\,668 YSO candidates in this region (these are YSO candidates located between 300 and 600 pc distance selected without applying any criterion based on parallax) with the SIMBAD database, and found 1\,701 pairs using 5$^{\prime\prime}$ search radius. We found that 1\,212 (71.3\%) sources are either YSOs, YSO candidates, T Tau stars, pre-main sequence stars, or Orion-type variables. Additional 412 sources (24.2\%) are stars in clusters, stars in nebulae, emission line stars, flaring stars, infrared sources and objects indicated as variable or irregular variable stars. These can also be interpreted as objects having properties similar to those of potential YSOs. 53 objects (3.1\%) are listed simply as ``star'' and 24 (1.4\%) were found to be classified by SIMBAD as some other type of object. Only one of the sources is known to be a BY Dra type object which is a late-type star, and no extragalactic object was found.

\subsection{YSO candidates among the \textit{Gaia} Photometric Science Alerts}

As \textit{Gaia} is monitoring the whole sky and collecting data, it detects transients and sudden brightness changes and is able to provide alerts to the astronomical community in almost real-time \citep{wyrzykowski2012}. The alerting system runs on a daily basis at the Institute  of  Astronomy  in Cambridge (part of the \textit{Gaia} Data Processing and Analysis Consortium, DPAC). The Nominal Scanning Law defines the observing strategy across the sky. It is a pattern optimised for the final astrometric solution \citep{lindegren2012}, and ensures that most of the stars will obtain, on average, about 70 measurements at different scanning angles. However, some areas of high stellar density such as the Galactic Bulge will only have about 50 measurements during the entire mission \citep{wyrzykowski2012}. 

The $\sim$70 brightness values are enough to be sure that \textit{Gaia} detects numerous very rare events like outbursts of FUor or EXor-type young stars, of which only a handful is known \citep{aduard2014}. Identification of potential YSO candidates can help the decision tree of the pipeline to promote a target with possible outbursts and fadings due to the circumstellar disk. Whenever an alert is made for such an object, the data can be made available, allowing for continuous follow-up of the candidate YSO. As of 2019 April 8$^{th}$ the \href{http://gsaweb.ast.cam.ac.uk/alerts/home}{\textit{Gaia} Photometric Science Alerts} database includes 7\,670 objects. 131 of them are known to be of YSO nature. %, as it is indicated in their classification or described in their Comment column.

%\subsubsection{YSO candidates among the Gaia Science Alerts}
We used a 1$^{\prime\prime}$ radius to match the alerts database to our Initial Dataset (Section~\ref{initialdata}). As a result we found 187 matching sources. The reason for finding only 187 objects is described below: 
\begin{enumerate}
\item The alerts are dominated by new sources, that were previously unseen and DR2 contains data collected from the first 22 months of the mission. As a result, only 1\,959 out of all the alerts have a DR2 counterpart.
\item Only 1\,012 of them had AllWISE counterparts.
\item As explained in Section~\ref{initialdata} we classified only those sources that are located in dusty environments and the \textit{Planck} dust opacity value is above 1.3$\times 10^{-5}$ and they fulfill our quality criteria (i.e. have photometric data in all bands and only one \textit{Gaia} source belongs to a given AllWISE ID), as well. Only 251 sources have met our requirements. 
\end{enumerate}

84 of the 251 matching sources were classified as YSO candidates with $P_R>0.5$ and $PL_Y\geq0.9$, or $P_R\leq0.5$ and $PS_Y\geq0.9$. In Table~\ref{tab:classifiedysoalerts} we listed their main properties. In case the nature of the alerted object is known, it is listed on the Alerts Index webpage. From the 84 object we classified as YSO candidates 44 were already known to be confirmed YSOs or YSO candidates. The other 40 sources had unknown object classes. This means that none of the alerts we classified as YSO candidates were known to be any different kind of object, e.g. Mira, cataclysmic variable, supernova, etc.

\clearpage
\onecolumn
\begin{center}
\begin{longtable}{|l|l|l|l|l|l|l|l|l|}
\caption{\textit{Gaia} Science Alerts classified as YSO candidates with $PL_Y\geq0.9$ and $P_R>0.5$ or with $PS_Y\geq0.9$ and $P_R\leq0.5$ according to our method. The columns are as follows: 1. Alert name as issued by the Alerts Index webpage, 2. Date and time of the alerting observation, 3.-4. RA and Dec, 5. object class based on literature, 6. probability of being a YSO according to our classification when WISE W3 and W4 photometry is included, 7. probability of being a YSO according to our classification when WISE W3 and W4 photometry is not used in the classification, 8. the probability of the \textit{WISE} WISE W3 and W4 photometry being real, 9. \textit{Gaia} DR2 source ID}	\label{tab:classifiedysoalerts}\\

  \hline 
  \multicolumn{1}{|c|}{Name} &
  \multicolumn{1}{c|}{Date} &
  \multicolumn{1}{c|}{RA} &
  \multicolumn{1}{c|}{Dec} &
  \multicolumn{1}{c|}{Class} &
  \multicolumn{1}{c|}{$PL_Y$} &
  \multicolumn{1}{c|}{$PS_Y$} &
  \multicolumn{1}{c|}{$P_R$} &
  \multicolumn{1}{c|}{$Gaia$ DR2 ID} \\
  \hline
\endfirsthead

\multicolumn{9}{c}%
{{\bfseries \tablename\ \thetable{} -- continued from previous page}} \\
\hline \multicolumn{1}{|c|}{Name} &
  \multicolumn{1}{c|}{Date} &
  \multicolumn{1}{c|}{Ra} &
  \multicolumn{1}{c|}{Dec} &
  \multicolumn{1}{c|}{Class} &
  \multicolumn{1}{c|}{$PL_Y$} &
  \multicolumn{1}{c|}{$PS_Y$} &
  \multicolumn{1}{c|}{$P_R$} &
  \multicolumn{1}{c|}{$Gaia$ DR2 ID} \\ 
  \hline 
\endhead

\hline \multicolumn{9}{|r|}{{Continued on next page}} \\ \hline
\endfoot

\hline \hline
\endlastfoot

  Gaia19bez & 2019-04-01 23:49:11 & 83.70199 & -5.70783 & YSO & 0.9998 & 0.9994 & 0.47 & 3017244319631273856\\
  Gaia19bdw & 2019-03-29 22:34:24 & 161.67619 & -60.16919 & YSO & 0.9552 & 0.94 & 0.492 & 5350284122620723072\\
  Gaia19azy & 2019-03-15 21:39:49 & 244.35307 & -36.95937 & YSO & 0.991 & 0.8764 & 0.65 & 6021662385163163648\\
  Gaia19axm & 2019-03-07 03:27:42 & 275.5886 & -15.80299 & unknown & 0.9918 & 0.9874 & 0.35 & 4097852452014666368\\
  Gaia19axl & 2019-02-08 12:15:35 & 83.85629 & -5.76245 & YSO & 0.9984 & 0.996 & 0.632 & 3017245144265120384\\
  Gaia19axj & 2019-02-15 12:39:42 & 87.26549 & 22.45243 & unknown & 0.9052 & 0.82 & 0.592 & 3427267972452130432\\
  Gaia19awc & 2019-03-05 03:14:14 & 273.90055 & -2.25697 & unknown & 0.8174 & 0.9072 & 0.19 & 4270884761539465344\\
  Gaia19avm & 2019-03-04 06:05:39 & 89.31937 & -9.71695 & unknown & 0.791 & 0.9264 & 0.224 & 3011393165424898176\\
  Gaia19avc & 2019-03-04 20:16:36 & 100.20506 & 9.74441 & YSO & 0.9828 & 0.9872 & 0.612 & 3326712411412675968\\
  Gaia19avb & 2019-02-08 18:08:24 & 90.94027 & -9.6506 & YSO & 0.9638 & 0.9256 & 0.55 & 3005362103626941312\\
  Gaia19ars & 2019-02-17 13:34:44 & 24.26503 & 65.00039 & unknown & 0.997 & 0.9978 & 0.51 & 513080267819937920\\
  Gaia19arj & 2019-02-19 13:34:26 & 38.12432 & 72.60243 & YSO & 0.9892 & 0.9824 & 0.538 & 545882891555329536\\
  Gaia19aqr & 2019-02-18 08:26:58 & 302.77041 & 33.91837 & unknown & 0.9068 & 0.9242 & 0.594 & 2055635703663536640\\
  Gaia19aqi & 2019-01-17 19:00:23 & 162.34187 & -59.18257 & YSO & 0.8814 & 0.928 & 0.428 & 5350433523059602304\\
  Gaia19apz & 2019-02-16 21:26:00 & 274.24053 & -19.89191 & unknown & 0.9776 & 0.996 & 0.356 & 4094447401937531648\\
  Gaia19apd & 2019-02-14 09:59:19 & 242.09364 & -39.0797 & YSO & 0.9994 & 0.9994 & 0.704 & 5997082867132347136\\
  Gaia19apa & 2019-02-12 18:32:03 & 83.21928 & 12.91914 & unknown & 0.9546 & 0.9382 & 0.47 & 3341084265336119552\\
  Gaia19ant & 2019-02-10 06:10:16 & 93.21496 & -6.18144 & YSO & 0.986 & 0.995 & 0.56 & 3019875657114864384\\
  Gaia19anj & 2019-02-09 00:16:26 & 84.51622 & -4.27858 & YSO & 0.9964 & 0.9728 & 0.516 & 3215624048669350016\\
  Gaia19anf & 2019-02-08 06:14:49 & 84.12543 & -6.38611 & YSO & 0.9998 & 0.9996 & 0.546 & 3016949616153731584\\
  Gaia19amy & 2019-02-08 06:14:38 & 84.03455 & -6.81008 & YSO & 0.9974 & 0.9978 & 0.562 & 3016920066778758528\\
  Gaia19ajk & 2019-01-29 22:04:15 & 300.39781 & 35.72513 & YSO & 0.9174 & 0.9012 & 0.398 & 2059471349978080256\\
  Gaia19aik & 2019-01-27 04:01:31 & 303.68998 & 40.25742 & unknown & 0.906 & 0.9194 & 0.688 & 2062414368951680000\\
  Gaia19ace & 2019-01-05 20:20:26 & 59.0241 & 53.6004 & YSO & 0.8828 & 0.9812 & 0.386 & 252155434006298112\\
  Gaia18eay & 2018-12-29 02:16:38 & 43.59314 & 58.61324 & unknown & 0.9504 & 0.9496 & 0.518 & 461362844316408704\\
  Gaia18ead & 2018-12-27 08:25:57 & 22.44645 & 68.47463 & unknown & 0.9342 & 0.9242 & 0.592 & 532063610946188416\\
  Gaia18dvz & 2018-12-18 01:28:32 & 129.72988 & -40.68815 & YSO & 0.9768 & 0.9734 & 0.568 & 5528238468964304512\\
  Gaia18dvs & 2018-12-14 08:38:32 & 302.93063 & 27.82686 & unknown & 0.9844 & 0.9762 & 0.514 & 1836621822164584832\\
  Gaia18drz & 2018-12-04 08:37:50 & 285.61527 & 9.02441 & YSO & 0.9966 & 0.997 & 0.538 & 4310567510575660928\\
  Gaia18dlu & 2018-11-17 09:06:29 & 304.4854 & 41.71411 & unknown & 0.9864 & 0.9868 & 0.458 & 2068593448191217920\\
  Gaia18dli & 2018-11-16 06:12:45 & 135.80023 & -47.82651 & unknown & 0.9828 & 0.968 & 0.464 & 5327027733703518720\\
  Gaia18dhk & 2018-11-04 18:36:05 & 199.71135 & -63.98177 & unknown & 0.938 & 0.9668 & 0.542 & 5859151160662602752\\
  Gaia18dhf & 2018-11-05 00:35:50 & 200.24079 & -63.24464 & unknown & 0.9674 & 0.97 & 0.492 & 5868194094041412096\\
  Gaia18dgx & 2018-11-03 12:31:44 & 189.92875 & -62.96378 & YSO & 0.9702 & 0.9708 & 0.552 & 6055161579577825280\\
  Gaia18dag & 2018-10-13 02:57:42 & 284.53971 & 3.01279 & unknown & 0.9856 & 0.996 & 0.428 & 4280720889482000128\\
  Gaia18czn & 2018-10-10 18:17:19 & 84.30843 & -7.06144 & YSO & 1.0 & 0.9994 & 0.534 & 3016853443246145792\\
  Gaia18czm & 2018-10-10 13:43:35 & 98.08678 & 5.07122 & YSO & 0.9938 & 0.989 & 0.332 & 3131338717100881024\\
  Gaia18cvo & 2018-09-30 08:25:28 & 256.4861 & -46.40277 & unknown & 0.9786 & 0.9772 & 0.62 & 5962994536340143360\\
  Gaia18cvc & 2018-09-28 17:29:57 & 67.07812 & 48.57304 & unknown & 0.9908 & 0.9938 & 0.362 & 258093106034362368\\
  Gaia18cqd & 2018-09-13 00:00:51 & 49.9249 & 11.50455 & unknown & 0.7004 & 0.903 & 0.456 & 17030889653951616\\
  Gaia18cpz & 2018-09-12 07:20:32 & 129.11882 & -39.2163 & unknown & 0.9992 & 0.9986 & 0.466 & 5528691566536872320\\
  Gaia18cpd & 2018-09-09 22:17:33 & 301.45445 & 28.84617 & unknown & 0.9518 & 0.9388 & 0.66 & 2029135824162001536\\
  Gaia18cpb & 2018-09-09 22:13:57 & 298.16966 & 27.10213 & unknown & 0.9456 & 0.9604 & 0.464 & 2027177392100413440\\
  Gaia18cnc & 2018-09-02 23:36:02 & 62.38696 & 48.02058 & unknown & 0.9844 & 0.9698 & 0.234 & 246335718959842688\\
  Gaia18cls & 2018-08-30 07:23:50 & 135.70595 & -46.74502 & unknown & 0.8928 & 0.9226 & 0.426 & 5330182610521277440\\
  Gaia18cgq & 2018-08-16 03:09:08 & 271.04164 & -24.32494 & YSO & 0.9854 & 0.9838 & 0.136 & 4065976338621931136\\
  Gaia18ccp & 2018-08-10 08:59:12 & 257.91176 & -36.03637 & unknown & 0.9526 & 0.9438 & 0.636 & 5977001696005980672\\
  Gaia18cbh & 2018-08-06 09:05:10 & 236.374 & -34.39422 & YSO & 0.9524 & 0.8986 & 0.554 & 6014692477866901760\\
  Gaia18cac & 2018-08-03 08:46:10 & 239.26591 & -54.15523 & YSO & 0.986 & 0.9708 & 0.462 & 5884723980058860544\\
  Gaia18byl & 2018-07-30 17:40:01 & 37.72747 & 59.5787 & unknown & 0.9414 & 0.922 & 0.546 & 465085344014342400\\
  Gaia18bxx & 2018-07-28 11:26:18 & 37.33175 & 73.03989 & YSO & 0.9758 & 0.9858 & 0.576 & 546658352195182464\\
  Gaia18bwj & 2018-07-23 23:39:00 & 343.37586 & 62.63983 & YSO & 0.964 & 0.9222 & 0.562 & 2207276292913481856\\
  Gaia18bvw & 2018-07-22 02:38:35 & 149.54376 & -58.17488 & unknown & 0.9706 & 0.986 & 0.544 & 5259160275404381440\\
  Gaia18bre & 2018-06-26 23:47:00 & 313.22393 & 44.24919 & YSO & 0.9354 & 0.9422 & 0.568 & 2162939208074527616\\
  Gaia18bjd & 2018-05-27 19:36:51 & 15.81917 & 61.61695 & YSO & 0.998 & 0.9954 & 0.57 & 522714223060690432\\
  Gaia18bgs & 2018-05-17 09:53:56 & 106.14958 & -11.08006 & YSO & 0.9916 & 0.9626 & 0.546 & 3046430340396482176\\
  Gaia18bfz & 2018-05-14 22:05:55 & 98.53708 & 4.49401 & unknown & 0.9978 & 0.999 & 0.458 & 3130544384371982336\\
  Gaia18beu & 2018-05-09 09:55:25 & 88.6581 & 1.49764 & YSO & 0.9996 & 0.9984 & 0.652 & 3315789764117572224\\
  Gaia18bdn & 2018-05-04 15:42:22 & 83.49683 & -5.7731 & YSO & 0.9208 & 0.9362 & 0.598 & 3017253871638555520\\
  Gaia18avw & 2018-04-03 23:56:14 & 39.71863 & 66.43154 & unknown & 0.9524 & 0.9418 & 0.512 & 516596707858306944\\
  Gaia18asj & 2018-03-23 22:55:50 & 79.91252 & 7.75051 & YSO & 0.9336 & 0.885 & 0.626 & 3241408897713545472\\
  Gaia18arc & 2018-03-19 15:44:50 & 134.72185 & -49.22867 & YSO & 0.9418 & 0.9764 & 0.376 & 5325391385523762048\\
  Gaia18afm & 2018-01-19 14:55:30 & 298.60582 & 28.07414 & unknown & 0.9334 & 0.9972 & 0.432 & 2028360878957519744\\
  Gaia17ccn & 2017-08-20 11:32:05 & 101.18753 & 0.22775 & unknown & 0.9296 & 0.9148 & 0.442 & 3125506215936019072\\
  Gaia17byj & 2017-08-02 04:40:04 & 82.3732 & 37.09014 & unknown & 0.9922 & 0.9858 & 0.534 & 184279626286154496\\
  Gaia17bxn & 2017-07-29 13:47:25 & 254.53228 & -40.96615 & unknown & 0.9522 & 0.9686 & 0.63 & 5966671582407739008\\
  Gaia17bse & 2017-07-04 16:44:46 & 305.20584 & 39.65007 & YSO & 0.9346 & 0.9704 & 0.28 & 2061491255911833088\\
  Gaia17bns & 2017-06-13 15:30:16 & 131.04029 & -41.33341 & YSO & 0.9946 & 0.997 & 0.394 & 5525226180763038336\\
  Gaia17bnf & 2017-06-11 16:49:37 & 320.56135 & 49.09411 & unknown & 0.9994 & 0.9988 & 0.55 & 2165017456849422080\\
  Gaia17aqb & 2017-03-07 19:16:04 & 314.37986 & 44.04373 & YSO & 0.9636 & 0.9654 & 0.548 & 2162145223244439168\\
  Gaia17afn & 2017-01-21 08:39:05 & 83.78629 & -4.7812 & YSO & 0.9706 & 0.9622 & 0.542 & 3209576459840251776\\
  Gaia17afi & 2017-01-20 08:37:02 & 83.94619 & -6.19588 & YSO & 0.9986 & 0.998 & 0.198 & 3017166907140903040\\
  Gaia17afh & 2017-01-20 08:35:56 & 84.72033 & -7.00669 & YSO & 0.9876 & 0.9948 & 0.57 & 3016107798267446528\\
  Gaia16bqh & 2016-10-26 19:07:28 & 68.21489 & 54.22959 & unknown & 0.9602 & 0.9554 & 0.512 & 273955008670915328\\
  Gaia16bpa & 2016-10-18 04:31:25 & 272.45945 & -21.76201 & unknown & 0.9912 & 0.9992 & 0.468 & 4069775842794238976\\
  Gaia16bmg & 2016-10-07 18:17:57 & 346.65814 & 61.00646 & unknown & 0.969 & 0.9876 & 0.56 & 2014871722373659776\\
  Gaia16blg & 2016-10-03 06:02:50 & 315.26076 & 52.45242 & YSO & 0.999 & 0.9994 & 0.598 & 2170293158455917440\\
  Gaia16bft & 2016-09-06 20:10:30 & 53.13751 & 31.03934 & YSO & 1.0 & 0.9992 & 0.536 & 121163093300288256\\
  Gaia16apz & 2016-05-22 14:20:39 & 343.5964 & 61.84157 & YSO & 0.8808 & 0.994 & 0.49 & 2206993787141103488\\
  Gaia16ama & 2016-04-26 04:01:06 & 344.98905 & 62.4256 & unknown & 0.9144 & 0.9988 & 0.498 & 2207037973764265472\\
  Gaia16aly & 2016-04-27 06:43:37 & 100.14171 & 9.80808 & YSO & 0.9746 & 0.986 & 0.148 & 3326716534581491840\\
  Gaia16alt & 2016-04-23 10:05:23 & 325.74996 & 66.19105 & YSO & 0.9966 & 0.9948 & 0.596 & 2218013543651547904\\
  Gaia16ajp & 2016-03-29 06:40:07 & 105.65842 & -11.49547 & unknown & 0.9942 & 0.988 & 0.182 & 3046035130386309888\\
  Gaia16agv & 2016-02-29 13:04:14 & 83.69805 & -5.96583 & YSO & 0.9946 & 0.9894 & 0.608 & 3017188931733323776\\
  Gaia16agu & 2016-02-29 13:02:36 & 85.10685 & -7.09369 & YSO & 0.9746 & 0.9604 & 0.64 & 3016110860580352128\\
\end{longtable}
\end{center}
\clearpage
\twocolumn

We also studied those 129 alerts that were already known to be YSOs, and checked how our classification performed in those cases. Previously we accepted a source as a YSO candidate if $P_R>0.5$ and $PL_Y\geq0.9$, or $P_R\leq0.5$ and $PS_Y\geq0.9$. Among these objects we found 15 that were not accepted as YSO candidates. We listed them in Table~\ref{tab:ysoalerts}.  Among these 15 cases we found only four cases where $PL_Y<0.5$ and also $PS_Y<0.5$:
\begin{itemize}
%\item Gaia18ajf ($PL_Y$=0.39, $PS_Y$=0.51) is a red variable star close to the Galactic plane and was found to be a YSO in our previous study in \citet{marton2016}
\item Gaia18ale ($PL_Y$=0.25, $PS_Y$=0.15) is also a red Galactic plane source that was classified as YSO in \citet{marton2016}
\item Gaia16aez ($PL_Y$=0.24, $PS_Y$=0.24) is a source known to the SIMBAD database as a YSO and was first listed as a young star by \citet{herbig2002}
\item Gaia19agd ($PL_Y$=0.14, $PS_Y$=0.09) is identified in the ASAS-SN system as an AGN or YSO, but is listed in multiple extragalactic catalogues in the VizieR database \citep[e.g.][]{liao2019}. 
\item Gaia18adm ($PL_Y$=0.06, $PS_Y$=0.09) is listed on the Alerts Index webpage as a possible YSO, but is classified as a possible AGN by \citet{edelson2012}.

\end{itemize}

\begin{table*}
	\small\addtolength{\tabcolsep}{-3pt}
    %\small
	\centering
	\caption{All \textit{Gaia} Science Alerts already known to be YSOs and found among our DR2 sources. Only 12 of them was classified as YSO candidate with $P_Y<0.9$ by our method, and only 4 of those have $P_Y<0.5$. The columns are as follows: 1) Alert name as issued by the Alerts Index webpage, 2) Date and time of the alerting observation, 3)-4) RA and Dec, 5)-6) probability of being a YSO according to our classification and its error, 7) \textit{Gaia} DR2 source ID}
	\label{tab:ysoalerts}
    
\begin{tabular}{|l|l|r|r|r|r|r|r|p{4.5cm}|}
\hline
  \multicolumn{1}{|c|}{Name} &
  \multicolumn{1}{c|}{Date} &
  \multicolumn{1}{c|}{RA} &
  \multicolumn{1}{c|}{Dec} &
  \multicolumn{1}{c|}{$PL_Y$} &
  \multicolumn{1}{c|}{$PS_Y$} &
  \multicolumn{1}{c|}{$P_R$} &
  \multicolumn{1}{c|}{$Gaia$ DR2 ID} &
  \multicolumn{1}{c|}{Comment} \\
\hline
  Gaia19ave & 2019-02-17 00:25:12 & 100.03513 & 9.99052 & 0.6802 & 0.6804 & 0.43 & 3326910568319185408 & known YSO 2MASS J06400842+0959259 fades by 1.6 mag\\
  Gaia19ape & 2019-02-14 10:54:41 & 277.46865 & -3.94831 & 0.665 & 0.6706 & 0.528 & 4257893967232850304 & candidate YSO dims by 0.6 mag\\
  Gaia19ajj & 2019-01-29 05:25:01 & 122.69077 & -36.07526 & 0.8866 & 0.8312 & 0.49 & 5544564391276237312 & Gaia star (possible YSO) brightens by 5 mags over 3 years\\
  Gaia19agd & 2019-01-19 02:22:19 & 327.40184 & 65.62713 & 0.1374 & 0.093 & 0.548 & 2219241148384022528 & YSO brightens by 1.5 mag\\
  Gaia18dnb & 2018-11-20 11:40:51 & 106.15048 & -11.26449 & 0.7266 & 0.7582 & 0.536 & 3046408758180843264 & candidate YSO brightens by 0.7 mag in \verb+~+1 month\\
  Gaia18bpt & 2018-06-20 16:52:58 & 182.60228 & -61.75732 & 0.8776 & 0.9068 & 0.61 & 6057768865245322240 & Gaia source coincident with candidate YSO shows gradual decline of almost 2 mags\\
  Gaia18bfo & 2018-05-14 10:14:45 & 98.66708 & 14.03232 & 0.559 & 0.4688 & 0.43 & 3355671456041735808 & YSO brightens by 2 mags\\
  Gaia18asa & 2018-03-21 06:29:28 & 328.05148 & 47.2505 & 0.7936 & 0.7824 & 0.28 & 1974730133391256832 & 1 mag fading of red Gaia source, previous dips in light curve. Possible YSO\\
  Gaia18apr & 2018-03-12 06:00:09 & 11.05205 & 55.15218 & 0.5646 & 0.582 & 0.594 & 418322186771625344 & erratic variable Gaia source brightens by 2.5 mags over 500 days, candidate YSO/AGN\\
  Gaia18ale & 2018-02-11 14:24:14 & 278.02124 & -20.84946 & 0.2486 & 0.1482 & 0.686 & 4091843208674333824 & 1.8 mag increase in variable red candidate YSO\\
  Gaia18ajf & 2018-02-05 08:27:11 & 261.90538 & -24.12577 & 0.3876 & 0.51 & 0.494 & 4111263818162632960 & variable red Galactic plane source brightens by >1 mag, candidate YSO (Marton et al.)\\
  Gaia18aiq & 2018-02-05 00:28:58 & 109.55913 & -23.72759 & 0.5744 & 0.3822 & 0.39 & 5617801486411760000 & 2 mag rise on previously variable Galactic Plane source, strong WISE signal, probably YSO\\
  Gaia18adm & 2018-01-13 10:10:38 & 126.44875 & -34.15915 & 0.0568 & 0.087 & 0.608 & 5543700827969368320 & possible YSO (showing dips), declines by 1 magnitude\\
  Gaia17bvo & 2017-07-23 07:38:32 & 238.2645 & -52.40214 & 0.8212 & 0.5944 & 0.248 & 5981188808079578624 & 0.5 mag brightening of Galactic plane variable source, candidate YSO\\
  Gaia16aez & 2016-02-19 16:32:26 & 328.37432 & 47.30115 & 0.242 & 0.2394 & 0.396 & 1974546686752048768 & sudden drop in flux in YSO 2MASS J21532984+4718041\\
\hline
\end{tabular}

\end{table*}

%\subsubsection{Unpublished alerts}
We also checked the database of the unpublished \textit{Gaia} alerts for sources we classified as YSO candidates. The database includes 6\,802\,636 sources as of 2019 April 9$^{th}$. Using a 1$^{\prime\prime}$ matching radius 6\,450 matching sources were found. This large number suggests that likely many more YSOs show brightness variations and shows the potential of the analysis of \textit{Gaia} light curves once they become public.

\section{Summary}

In this paper we presented a classification of those sources in the \textit{Gaia} DR2 catalogue that have a counterpart in the AllWISE catalogue. Robust samples of known main--sequence stars, evolved stars, extragalactic objects and YSOs were built to create clean dataset including \textit{Gaia}, 2MASS and \textit{WISE} photometry. We used these datasets as training samples for several machine learning algorithms and tested them to find the method best working for our purposes. 

We found that Random Forests with 500 trees and 5 random variables tried at each split gives the best results. For all 101\,838\,724 sources of our initial sample the classification was done, first by including longer-wavelength $W3$ and $W4$ \textit{WISE} photometric data, then by excluding them. Latter classification was necessary, because the $W3$ and $W4$ photometric data are unreliable in many cases in crowded regions, where YSOs are typically found. In order to decide if the $W3$ and $W4$ photometric data is reliable, we visually inspected $W3$ and $W4$ images of YSOs, created a Random Forest classifier and for all sources we calculated the probability of having reliable detections in those bands.

All our results, including the classification probabilities}, source IDs and photometric data are publicly available through the \href{http://vizier.u-strasbg.fr/}{VizieR} service.

Our careful validation process showed that we are able to recover $\sim$92\% of the known YSOs and the fraction of sources false positively classified as YSO candidates is around 6\%. Our method was successfully tested in a realistic scenario, the Orion A star forming region. The results are in excellent agreement with that of \citet{grossschedl2018}, confirming the 3D structure of Orion A.

We also used our classification to identify previously unknown YSO candidates in the \textit{Gaia} Photometric Science Alerts System and were able to add 40 more to the list of 130 YSO alerts, significantly increasing their number. A more detailed analysis of YSO light curves based on this classification and YSO identification will be helpful to improve our understanding of YSO variability, disk evolution and planet formation theories and models. Also, the alerting algorithms can benefit: the AlertPipe of \textit{Gaia} and future telescopes can be more sensitive to YSO events triggering more alerts for the detailed follow-up observations.

\section*{Acknowledgements}

First of all, we want to acknowledge the comments and suggestions of our anonymus referee, who really helped to improve the content and understanding of the paper.

G. Marton and this work was supported by the Hungarian National
Research, Development and Innovation Office (NKFIH)
grant PD-128360. This project has received funding from the European Research Council (ERC) under the European Union's Horizon 2020 research and innovation programme under grant agreement No 716155 (SACCRED). This work has made use of data from the European Space Agency (ESA) mission {\it \textit{Gaia}} (\url{https://www.cosmos.esa.int/gaia}), processed by the {\it \textit{Gaia}} Data Processing and Analysis Consortium (DPAC, \url{https://www.cosmos.esa.int/web/gaia/dpac/consortium}). Funding for the DPAC has been provided by national institutions, in particular the institutions participating in the {\it \textit{Gaia}} Multilateral Agreement. This publication makes use of data products from the Wide-field Infrared Survey Explorer, which is a joint project of the University of California, Los Angeles, and the Jet Propulsion Laboratory/California Institute of Technology, funded by the National Aeronautics and Space Administration. This research has made use of the SIMBAD database, operated at CDS, Strasbourg, France. This research has made use of the VizieR catalogue access tool, CDS, Strasbourg, France. The original description of the VizieR service was published in A\&AS 143, 23. Many figures and analyses were done with the TOPCAT software \citep{taylor2005}.

%%%%%%%%%%%%%%%%%%%%%%%%%%%%%%%%%%%%%%%%%%%%%%%%%%



%\bibliographystyle{mnras}


%%%%%%%%%%%%%%%%%%%%%%%%%%%%%%%%%%%%%%%%%%%%%%%%%%

%%%%%%%%%%%%%%%%% APPENDICES %%%%%%%%%%%%%%%%%%%%%
%
\appendix
\newpage
\section{Photometric YSO catalogues found in the \href{http://vizier.u-strasbg.fr/}{VizieR database}}\label{appendix:a}
\citet{allen2012}, \citet{billot2010}, \citet{connelley2008}, \citet{evans2009}, \citet{furlan2016}, \citet{gutermuth2008}, \citet{gutermuth2009}, \citet{jose2017}, \citet{kirk2009},  \citet{kun2016b}, \citet{lada2017},  \citet{rebull2011}, \citet{saral2017}

\section{Spectroscopic YSO catalogues listed by VizieR}\label{appendix:b}
\citet{alcala2014}, \citet{ansdell2016,ansdell2017}, \citet{connelley2010}, \citet{cooper2013}, \citet{dzib2015}, \citet{erickson2015}, \citet{fang2009}, \citet{frasca2017}, \citet{kim2016}, \citet{kumar2014}, \citet{kun2009}, \citet{kun2016}, \citet{oliveira2009}, \citet{pascucci2016}, \citet{rebollido2015}, \citet{szegedi2013}

\section{YSO related Spitzer publications}\label{appendix:c}
\citet{alcala2008}, \citet{allen2008}, \citet{an2011}, \citet{balog2007}, \citet{broekhoven2014}, \citet{bryden2009}, \citet{chavarria2008}, \citet{chen2011}, \citet{chen2012}, \citet{cieza2007}, \citet{cloutier2014}, \citet{cody2014}, \citet{dahm2007}, \citet{dewangan2011}, \citet{dunham2015}, \citet{flaherty2013}, \citet{harvey2007}, \citet{hernandez2007a}, \citet{hernandez2007b}, \citet{hernandez2008}, \citet{jose2016}, \citet{koenig2008}, \citet{kounkel2016}, \citet{lada2006}, \citet{lopez2013}, \citet{luhman2008}, \citet{luhman2010}, \citet{megeath2012}, \citet{merin2008}, \citet{muench2007}, \citet{oliveira2010}, \citet{oliveira2013}, \citet{peterson2011}, \citet{puga2009}, \citet{ragan2009}, \citet{rapson2014}, \citet{rebull2010}, \citet{rivera2011}, \citet{roccatagliata2011}, \citet{samal2012}, \citet{saral2015}, \citet{stelzer2009}, \citet{strafella2010}, \citet{vandermarel2016}, \citet{willis2013}, \citet{winston2007},  \citet{young2015}

\newpage
\section{Brightness -- $\tau$ and colour -- colour diagrams}\label{appendix:diagrams}
\begin{figure}
	\includegraphics[width=0.5\textwidth]{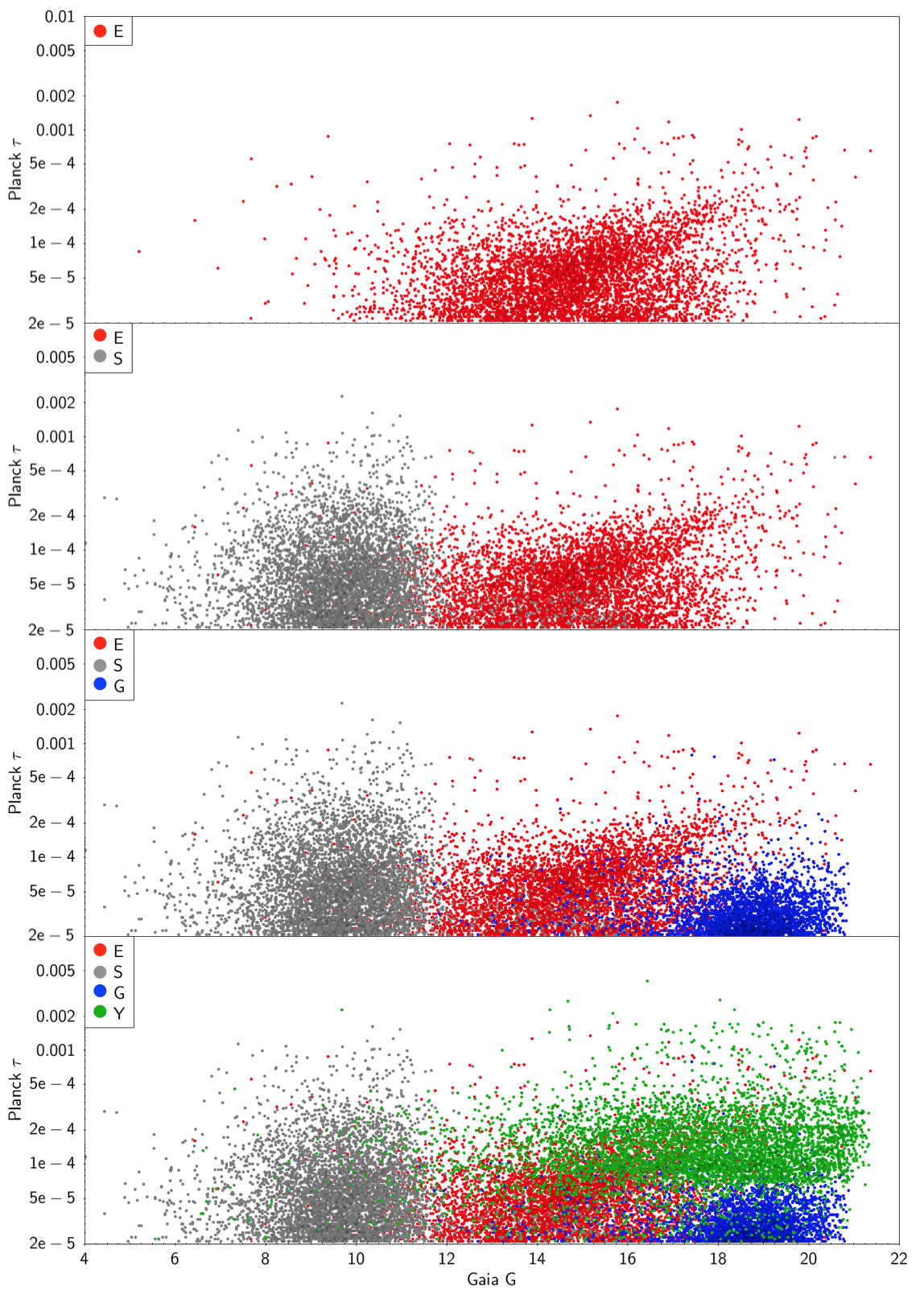}
    \caption{The 4 classes of the objects in the $G$ magnitude--$\tau$ diagram. The top panel shows the distribution of the evolved objects (E) with red dots. They are mostly located at intermediate \textit{Gaia} $G$ band values, between 12 and 18 magnitudes, and are at lower $\tau$ dust opacity values. On the second panel the main sequence stars (S) are plotted with gray dots. They apper to be brighter than the evolved objects, and a higher fraction is located at high $\tau$ values. On the third panel the extragalactic objects (G) are shown with blue dots and they are at the faint end of the diagram, with low $\tau$ values. On the bottom panel the YSOs (Y) are plotted with green dots. They are mostly faint and at higher $\tau$ values than the extragalactic objects, but lots of them are overlapping with the evolved stars.}
    
%    G (extragalactic, blue dots) tend to be fainter in the \textit{Gaia} $G$ band, while S (main sequence stars, gray dots) objects appear to be brighter. Also, more of the G sources are found at low $\tau$ values. Y (YSO, green) type objects are found at higher $\tau$ values, which is expected as star formation occurs in places with lots of ISM. E (evolved type) objects are mainly located between the Y and S sources.}
    \label{fig:bt}
\end{figure}

\begin{figure}
	\includegraphics[width=0.5\textwidth]{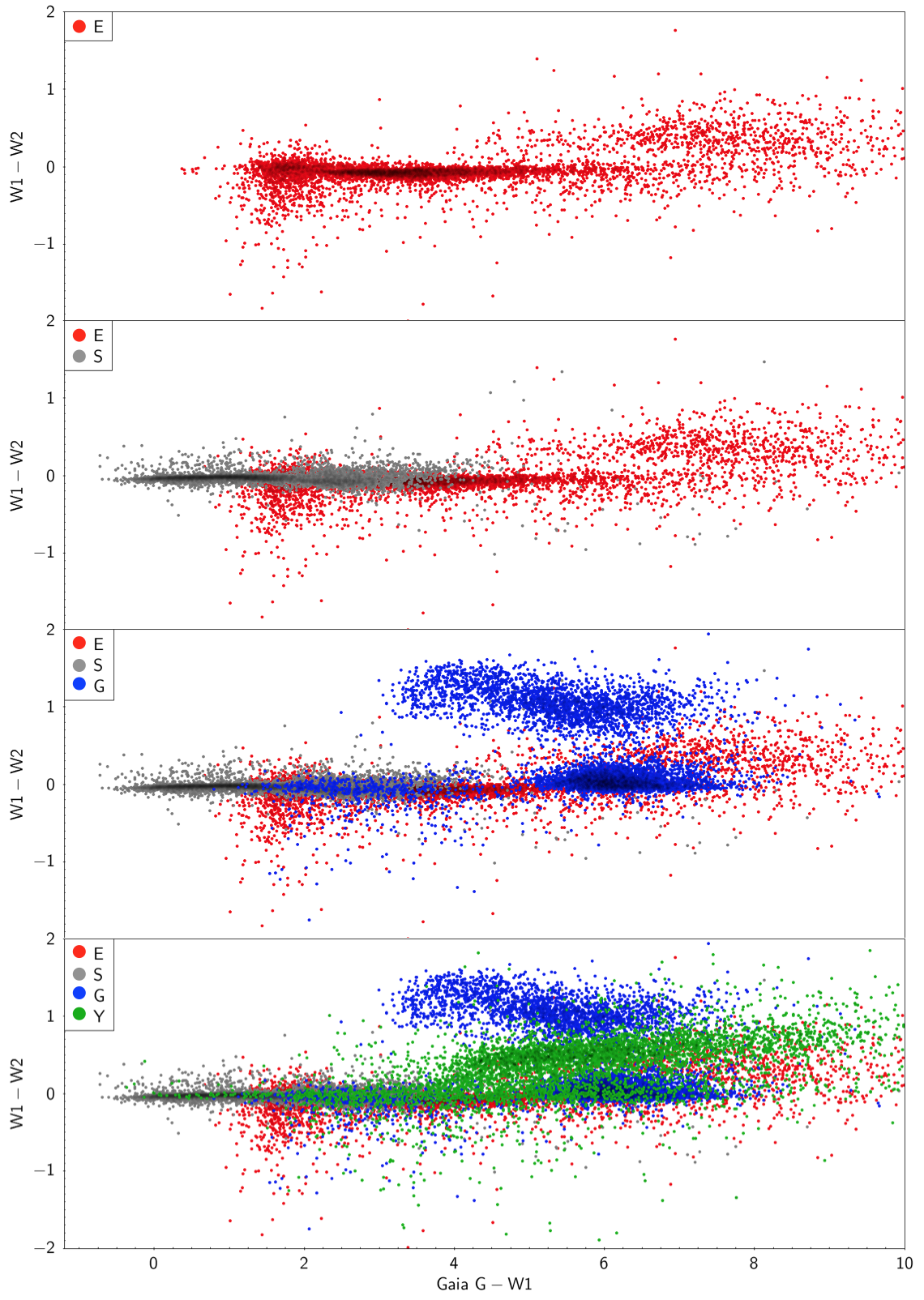}
    \caption{The colors represent the same type of objects as in Figure~\ref{fig:bt} on the $G-W1$ vs. $W1-W2$ colour-colour plane. Along the $G-W1$ axis the evolved stars are located from 1 to 10. Along the $W1-W2$ axis most of the sources are located around 0, but a smaller cloud of point is visible around 0.5 ranging from 6 to 10 in the $G-W1$ colour. These sources are mostly dominated by Mira type variables, while the others are mostly AGB, RGB and RR Lyr type objects. The main sequence stars have $G-W1$ colours spreading from -0.5 to 4, but their $W1-W2$ colours are very close to 0. In case of the extragalactic sources one can see a bimodial behavior, as well. The objects having $W1-W2$ colours close to 0 are mostly AGN and QSO type sources, while those with $W1-W2$ colours around one are simply indicated in SIMBAD as "Galaxy". The YSOs are mostly redder in both directions, although a bimodial distribution is present for them, too. One cluster of the sources has lower $G-W1$ color index and their $W1-W2$ colour is close to 0, highly overlapping with the main sequence stars. Another cloud of YSO dots is visible at higher $G-W1$ colours and they have a higher $W1-W2$ colour, as well, overlapping with the evolved stars and the extragalactic sources. }
    \label{fig:cc}
\end{figure}

\begin{figure}
	\includegraphics[width=0.5\textwidth]{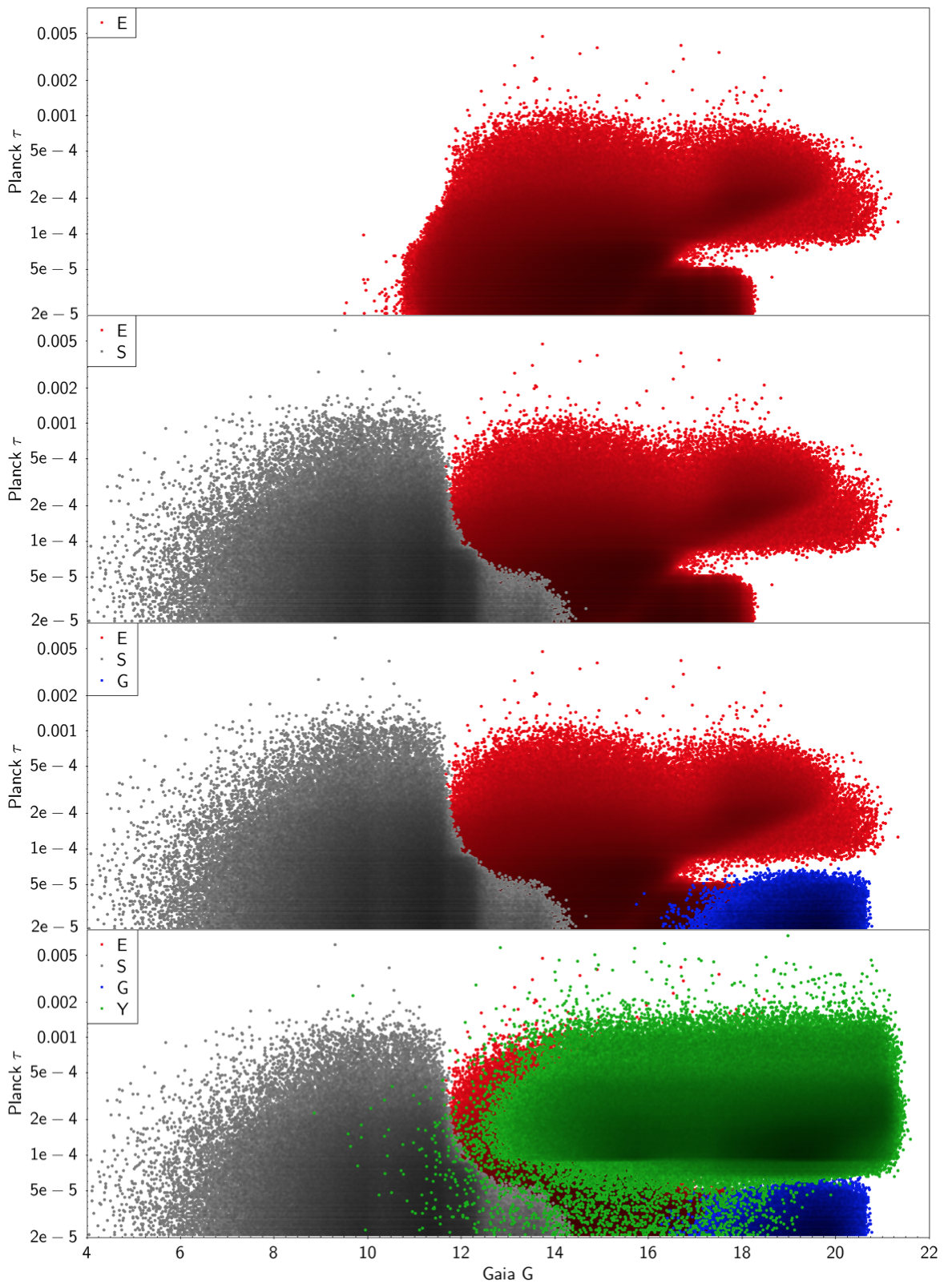}
    \caption{Position of the unknown sources in the same $G$ brightness--$\tau$ diagram as in Figure~\ref{fig:bt} after we classified them into the 4 object classes. Only those objects are plotted which were classified into the corresponding classis with $P\geq 0.9$. The position of the sources is very similar to those in Figure~\ref{fig:bt}.}
    \label{fig:classifiedbt}
\end{figure}

\begin{figure}
	\includegraphics[width=0.5\textwidth]{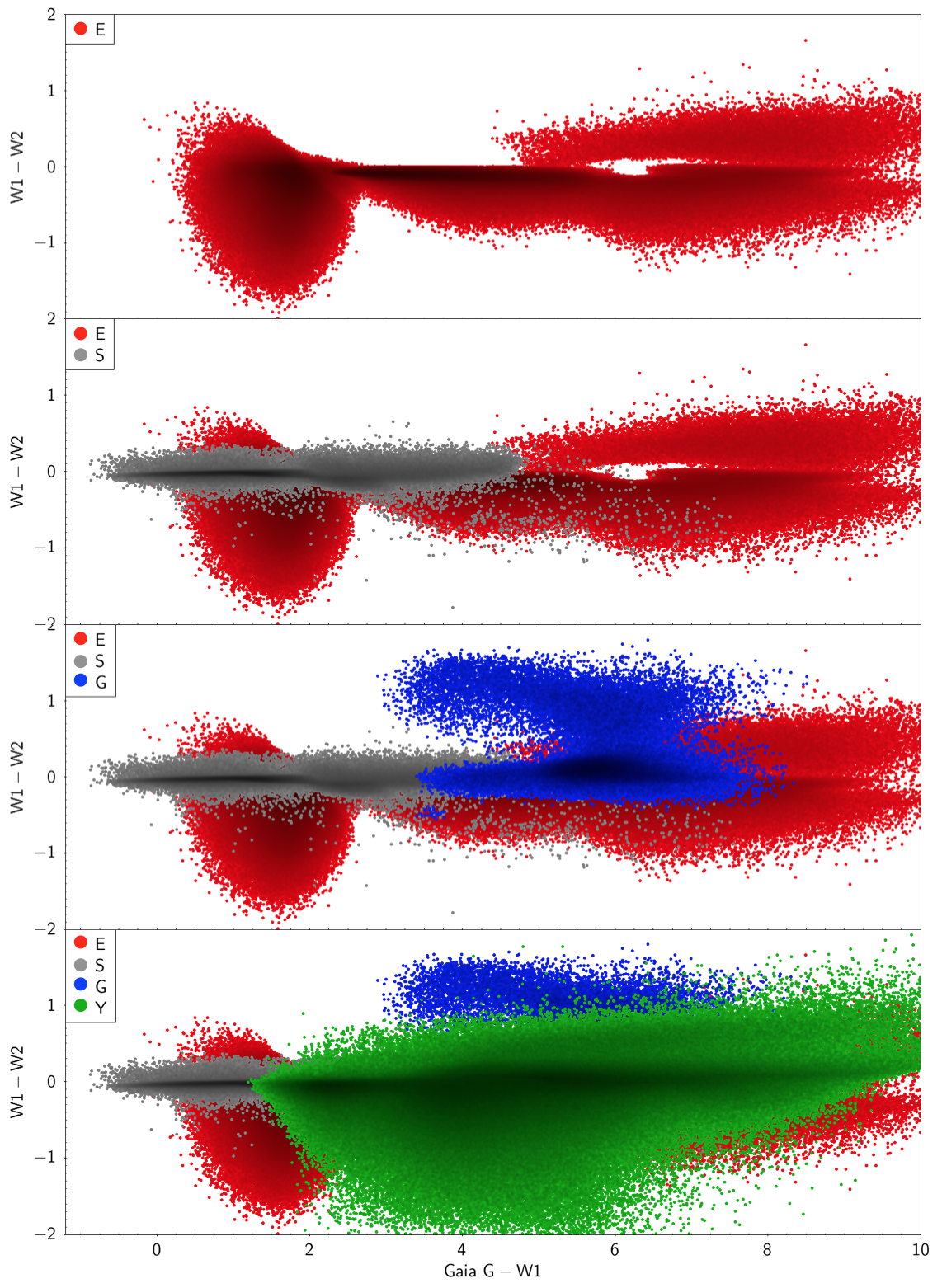}
    \caption{Position of the unknown sources in the same $G-W1$ vs. $W1-W2$ colour--colour diagrams as in Figure~\ref{fig:cc} after we classified them into the 4 object classes. As in Figure~\ref{fig:classifiedbt}, only those objects are plotted which were classified into the corresponding classis with $P\geq 0.9$. The position of the sources is very similar to those in Figure~\ref{fig:cc}, and the same substructures can be identified on the diagrams.}
    \label{fig:classifiedcc}
\end{figure}

%%%%%%%%%%%%%%%%%%%%%%%%%%%%%%%%%%%%%%%%%%%%%%%%%%

% Don't change these lines
\bsp	% typesetting comment
\label{lastpage}
\end{document}